\documentclass[default]{sn-jnl}

\usepackage{geometry}
\graphicspath{{./images/}{.}}

\usepackage{graphicx}%
\usepackage{multirow}%
\usepackage{amsmath,amssymb,amsfonts}%
\usepackage{amsthm}%
\usepackage{mathrsfs}%
\usepackage[title]{appendix}%
\usepackage{xcolor}%
\usepackage{textcomp}%
\usepackage{manyfoot}%
\usepackage{booktabs}%
\usepackage{algorithm}%
\usepackage{algorithmicx}%
\usepackage{algpseudocode}%
\usepackage{listings}%

\theoremstyle{thmstyleone}%
\theoremstyle{thmstyletwo}%

\theoremstyle{thmstylethree}%

\usepackage{physics}
\usepackage{subcaption}
\usepackage{tikz}

\newcommand{\fib}{\textbf{1}}
\newcommand{\vac}{\textbf{0}}

\begin{document}

\title{
Systematic Computation of Braid Generator Matrix in Topological Quantum Computing
}


\author*[1,2]{\fnm{Abdellah} \sur{Tounsi}}\email{abdellah.tounsi@umc.edu.dz}
\author[1,2]{\fnm{Nacer Eddine} \sur{Belaloui}}
\author[2,3]{\fnm{Mohamed Messaoud} \sur{Louamri}}
\author[1,2]{\fnm{Amani} \sur{Mimoun}}
\author[1,2]{\fnm{Achour} \sur{Benslama}}
\author*[1,2]{\fnm{Mohamed Taha} \sur{Rouabah}}\email{m.taha.rouabah@umc.edu.dz}

\affil*[1]{\orgdiv{\centering Constantine Quantum Technologies}, \orgname{\\Fr\`{e}res Mentouri University Constantine 1}, \orgaddress{\street{Ain El Bey Road}, \city{Constantine}, \postcode{25017}, \country{Algeria}}}

\affil[2]{\orgdiv{\centering Laboratoire de Physique Math\'{e}matique et Subatomique}, \orgname{\centering \\Fr\`{e}res Mentouri University Constantine 1}, \orgaddress{\street{Ain El Bey Road}, \city{Constantine}, \postcode{25017}, \country{Algeria}}}
\affil[3]{\orgdiv{Theoretical Physics Laboratory}, \orgname{University of
Science and Technology Houari Boumediene}, \orgaddress{\street{BP 32 Bab Ezzouar}, \city{Algiers}, \postcode{16111}, \country{Algeria}}}

\maketitle

\begin{abstract}
		We provide a comprehensive {systematic method }
		for the numerical computation of elementary braid operations in topological quantum computation (TQC). This {procedure} is systematically applicable to all anyon models, including $SU(2)_k$. Braiding non-abelian anyons is the essence of TQC, offering a topologically protected implementation of quantum gates. However, obtaining elementary braid matrix representations starting from the fusion and rotation matrices of a specific anyon model is {theoretically guarenteed but no numerical method is given, especially for systems with numerous anyons and complex fusion patterns. }
		Our proposed method addresses this challenge, first in the special case of sparse encoding, allowing for the inclusion of an arbitrary number of anyons per qudit, {and in the general case}. This is accomplished by introducing two methods, one is based on a novel braiding move we call knitting, {the second introduces more general algorithm which is optimal in number of required moves}. The method plays a key role in a broad topological quantum circuit simulator, enabling the examination and study of complex quantum circuits within the TQC framework. Importantly, it proves effective across various anyonic models, accommodating diverse fusion rules. We validate the method by simulating an approximated CNOT gate and present a first-of-a-kind GHZ state simulation on five qubits using three Fibonacci anyons per qubit.
\end{abstract}

	\keywords{Topological quantum computing, Braid group, Anyons, Entanglement}

	\maketitle
	
	\section{Introduction}
	A quantum computer leverages the fundamental principles of quantum mechanics to solve a class of computationally hard mathematical problems \cite{manin1980, Benioff1980, Feynman1982,Nielsen2010, Arute2019}.
	Instead of using classical Boolean bits, a quantum computer utilizes qubits, which are quantum states capable of being in superposition and entangled with each other. Quantum gates are unitary operators that act on the Hilbert space generated by qubits, thus allowing the manipulation of quantum information \cite{Nielsen2010}.
	Experimental realizations of quantum computers are vulnerable to errors that arise from decoherence, which is the result of the interaction of qubits with their environment \cite{Schlosshauer2005}.
	This phenomenon presents a significant challenge to the practical implementation of quantum computers. To address this issue, various approaches are under investigation, including the isolation of qubits from environmental disturbances and the use of cooling systems to minimize the effect of decoherence \cite{Khodjasteh2005, Morello2006, Zwanenburg2013}, as well as quantum error mitigation techniques \cite{cai2022quantum}. However, to fully harness the advantages of quantum computing, fault-tolerant  quantum error correction (QEC) is necessary. Nevertheless, the practical application of QEC is limited by the threshold theorem, which requires a sufficiently low error rate  \cite{doi:10.1137/S0097539799359385}.
	Furthermore,  the field of condensed matter physics has witnessed various experimental and theoretical discoveries in recent decades, revealing intriguing characteristics of the topological phase of matter \cite{Chen2013, Mesaros2013, Goldman2016, Halden2017, Wen2017, Zhang2018, Tasaki2018, Zeng2019}. 
	Hence, topological quantum systems are emerging as a promising platforms to store and process quantum information in a robust manner, through quantum evolutions that are immune to decoherence \cite{Dennis2002, Kitaev2003, Freedman2003, Pachos2012, Lahtinen2017, Stanescu2017, Field2018}. 
	Topological quantum computation (TQC) deals with two-dimensional quantum systems that support excitations with fractional statistics, known as anyons \cite{Wilczek1982, Khare2005}. These anyons exhibit statistics that differ from those of fermions and bosons. A system of $N$ non-abelian anyons possesses a topologically protected Hilbert space that grows exponentially with the number of anyons, and quantum information can be processed through the braiding of anyons \cite{Lerda1992, Kauffman2004, Bonesteel2005}. Non-abelian anyons promise to offer a fault-tolerant method of performing universal quantum computation, as information is stored in the topological, i.e., non-local features of the system, making the quantum state resilient, in principle, to conventional sources of decoherence \cite{Collins2006, Sarma2015, Field2018}.
	
	The execution of topological quantum computation can be divided into three main steps:
	\begin{enumerate}
		\item Initialization: pairs of non-abelian anyons are created from the vacuum. However, the creation of pairs is vulnerable to noise and not naturally robust. The distillation technique, which relies purely on braiding, can be employed to establish a suitable code space for initialization \cite{PhysRevA.81.052309}.
		\item Processing: the anyons are moved in 2-dimensional space to form braids in a 2+1-dimensional space-time, which corresponds to quantum unitary gates performed on the anyons' Hilbert space. The anyons must be kept sufficiently far apart to minimize errors. Nonetheless, a measurement-only scheme that does not require any exchange of anyons has also been proposed \cite{BONDERSON2009787}.
		\item Readout: the adjacent anyon pairs are fused together to close the computation process, which can be accomplished through the non-abelian anyons interferometry \cite{bonderson}.
	\end{enumerate}
	{According to the monoidal tensor category theory, the computation of braid matrix representation is guarenteed by knowing the fusion and rotation matrices of a given anyon model. However, the systematic procedure to determine the necessary sequence of fusion transformations is not studied. } 
	This work aims to present a comprehensive procedure for computing the matrix elements of braid operators, with a specific focus on applying the foundational principles of anyon model theory in quantum computing. In Section \ref{sec:anyons}, we provide a detailed description of the method for constructing qubits or qudits by selecting suitable fusion states of a set of identical anyons. In Section \ref{sec:calculating-breading-generators}, we present a generalized approach for systematically calculating the matrix components of braid operators using predefined fusion and rotation matrices. Within this section, we introduce a braiding move that we call knitting, which greatly simplifies the numerical computation of braid operators acting on two anyons from different qubits. Additionally, we analyze the computational complexity of this formula using concrete examples from the Fibonacci anyon model. {Furthermore, we provide a general algorithm that automates the process of computing braid generators in any arbitrary fusion case.} 
	Consequently, in Sections \ref{sec:cnot} and \ref{sec:ghz}, we demonstrate the utility of this method in tackling genuine quantum circuits using anyons. In Section \ref{sec:cnot}, the method is employed to reproduce a previously approximated CNOT gate using 6 Fibonacci anyons, while in Section \ref{sec:ghz}, we study the preparation of a five-qubit GHZ state using 15 Fibonacci anyons. It's noteworthy that the outcomes of this study have been implemented to develop an open-source package called TQSim \cite{TQSim}.
	
	\section{{Quantum Computation with} Anyons}
	\label{sec:anyons}
	
	Particles exhibit a unique statistical behavior when existing in two-dimensional space. Unlike in three-dimensional space where particles behave as either bosons or fermions, the statistical behavior of indistinguishable particles in two-dimensional space is not restricted to these two categories. The exchange of indistinguishable particles in three dimensions is governed by the permutation group, where the paths of exchanging two particles are irrelevant. Namely, the exchange of fermions or bosons results in a scalar phase factor $e^{i\phi}$ with $\phi = 0(\pi)$ for bosons(fermions).
	However, in two dimensions, the path of exchanging two particles falls into different homotopy classes, thereby allowing for particles of any statistical behavior called anyons \cite{leinaas1977, Wilczek1982, Lerda1992}. The statistical quantum evolution of anyons is described by the braid group, which is significant in TQC as the braiding operation is impervious to dynamical and geometrical perturbations.
	Two types of anyons can be distinguished: abelian and non-abelian anyons. Abelian anyons have abelian phase factor and give rise to a non-degenerate state, whereas non-abelian anyons have non-abelian phase factor and construct a degenerate quantum state \cite{Pachos2012, bhattacharjee2017}. This degeneracy is crucial for encoding non-trivial quantum states.
	In the context of TQC, quantum information is encoded in the topology of multiple anyons' state, rather than being encoded in the dynamical properties of a system, which can be easily altered by external perturbations \cite{Pachos2012}. As a result, the only operations that modify the quantum states implemented in TQC are non-local operations, which consist of braiding the world lines of anyonic particles. This approach offers robustness in the implementation of quantum computation, as the movement of particles around each other in an undesired way is less likely to occur when the particles are far apart \cite{Hormozi2007}. Therefore, anyons present an opportunity to perform reliable quantum computation without the need for extensive error correction overhead. Additionally, they provide a platform for constructing new quantum algorithms or reformulating existing algorithms in the TQC framework. One such breakthrough is the Aharonov-Jones-Landau (AJL) algorithm, which can approximate the Jones polynomials \cite{Aharonov2009, Field2018}, known to be a BQP-complete problem \cite{Greg2009}.
	Practically, anyons can exist as quasi-particles in effective two-dimensional surfaces. The creation of anyons can be accomplished experimentally by exploiting topological superconducting nanowire devices \cite{Mourik2012, PhysRevB.107.245423} or the fractional quantum Hall effect (FQHE) \cite{chakraborty1995quantum, Yoshioka1998, Bravyi2006, STERN2008204, Willett2013}. It has been confirmed that the quasi-particles produced by this latter exhibit anyonic statistics \cite{Bartolomei2020, Nakamura2020}. 
	Moreover, lattice models such as Kitaev toric code \cite{Kitaev2003}, Kitaev honeycomb model \cite{Kitaev2006}, and Levin-Wen model \cite{Levin2005, PRXQuantum.3.040315} are among the other potential candidates for the emergence of non-abelian anyons, a crucial element for the construction of topological quantum computer.\\
	
	Anyon models serve as the frameworks for TQC, and the appropriate mathematical language to discuss them is the Unitary Modular Category (UMC) theory. In this latter, the Jones-Wenzl projectors are utilized to model anyons \cite{PRXQuantum.2.010334}. Abelian anyons can be classified based on their braiding statistics, which measure their interactions when they are braided around one another. On the other hand, non-Abelian anyons can be further classified by their fusion statistics, which measure how they combine to form new anyons. 
	Consequently, the fusion statistics of non-Abelian anyons can be utilized {to encode} information, and it becomes possible to implement a universal set of quantum gates by braiding them around one another. Specifically, the $SU(2)_k$ anyon models for $k>2$ have been shown to be Turing-complete or universal \cite{Freedman2003}.
	The rules for combining two anyons into a larger composite are called fusion rules. For non-abelian anyons, different fusion channels are possible, which determine the fusion rules between different types of anyons denoted as $a$ and $b$, each associated with distinct charges. The allowed fusion outcomes are determined by the following rule:
	\begin{equation}
		\label{eq:fusion_rules}
		a \times b = \sum_i N_{ab}^{i} \ {i} .
	\end{equation}
	The value of the integer $N_{ab}^i$ refers to the count of distinct ways to generate the anyon $i$ by fusing together the anyons $a$ and $b$. The vacuum, represented as $\vac$, fuses trivially with any other anyon such that $a \times \vac = a$ for all anyons $a$. Consequently, an anti-anyon $\bar{a}$ can be defined for each anyon $a$, where $a \times \bar{a} = \vac$.
	The fusion processes comprise a collection of states that are orthogonal to each other, spanning a Hilbert space denoted as $\mathcal{F}$ and referred to as the fusion space. Each state in $\mathcal{F}$ specifies the outcomes of the fusion processes of multiple anyons following a specific order. 
	In the context of anyon models, the quantum information is encoded in the feasible fusion states of the anyons and manipulated by the exchange of anyons, a process known as braiding.
	Importantly, the fusion state of a pair of anyons is a non-local collective property of that pair which is not accessible by local observations on either anyon.
	As such, the quantum information must be resilient to local perturbations.
	Let us consider the fusion space of four anyons described by the following fusion states:
	\begin{equation}
		\label{state}
		\ket{(((a, b)_i,c)_j,d)_k} ,
	\end{equation}
	where the indexed parenthesis represents the fusion process. In this example, the fusion order is from left to right. This notation is interchangeable with the diagrammatic fusion tree representation, with the exception of a normalization factor \cite{bonderson}. The state \eqref{state} can also be represented as a tensor product of all associated fusion processes:
	\begin{equation}
		\label{eq:state-tensor}
		\ket{(((a, b)_i,c)_j,d)_k} = \ket{(a,b)_i}\otimes\ket{(i, c)_j}\otimes\ket{(j, d)_k} .
	\end{equation}
	Thus, the dimension of the fusion space, $\mathcal{F}_4$, of four-anyon system can be expressed as the summation of all possible ways of fusing anyons $a$, $b$, $c$ and $d$ through the formula
	\begin{equation}
		\dim(\mathcal{F}_4) = \sum_{ijk} N_{ab}^i N_{ic}^j N_{jd}^k  \ , 
	\end{equation}
	where $N_{ab}^i$ represents the number of distinct fusion channels of anyons $a$ and $b$ resulting in the anyon $i$. In the subsequent sections, only multiplicity-free anyon models, which satisfy $N_{ab}^i = 0$ or $1$ for all $a$, $b$, and $i$, will be considered. It is worth noting that most anyon models encountered in physical contexts fall under this case. \\
	
	The fusion of more than two anyons is not uniquely determined by any specific order of pairwise fusion processes due to the variability of intermediate quantum states that result from different fusion process orders. To account for the various possible ways of fusing more than two anyons, it is necessary to introduce fusion matrices, which enable basis transformations. Specifically, the fusion matrices $F_{abc}^j$ are defined as linear transformations that relate the only two ways of ordering the fusion of anyons $a$, $b$, and $c$ to the resulting anyon $j$, as follows:
	\begin{equation}
		\label{eq:F_matrix_def}
		\ket{((a, b)_i,c)_j} = \sum_k  (F_{a b c}^j)_k^i \ket{(a, (b,c)_k)_j}.
	\end{equation}
	The index $k$ corresponds to the intermediate charges that are summed over, and the coefficients $(F_{abc}^j)_k^i$ determine the amplitudes of the possible fusion outcomes. It is essential for these fusion matrices to be unitary to maintain the normalization of fusion states \cite{bonderson, Kitaev2006, Pachos2012, Stanescu2017}.
	On the other hand, braiding two anyons that fuse in a particular channel cannot change the fusion channel. This is because the total topological charge of the pair is not a local property, and it does not depend on the evolution of the pair. 
	The phase factor $R^i_{ab}$ is a property of the anyons $a$, $b$, and $i$. It is a measure of how the anyons interact when they are braided around each other. 
	The value of the phase factor $R^i_{ab}$ depends on the type of anyons that are being braided. Therefore, the braiding matrix $R_{ab}$ is defined by its operation on the state $\ket{(a,b)_i}$ as follows \cite{bonderson}:
	\begin{equation}
		\label{eq:R_matrix_def}
		R_{ab} \ket{(a,b)_i} = R_{ab}^{i} \ket{(b,a)_i}.
	\end{equation}
	To ensure consistency in an anyon model, the $F$ and $R$ matrices should satisfy the hexagon and pentagon equations, as demonstrated in any braided monoidal category \cite{Bakalov2000}.
	Furthermore, by solving the hexagon and pentagon identities, one can determine the numerical values of the components of the $F$ and $R$ matrices \cite{bonderson}. Another method of determining the components of those matrices for an $SU(2)_k$ anyon model with an arbitrary integer $k$, representing the level of the theory, is mentioned in \cite{PRXQuantum.2.010334, bonderson}. The algebraic theory of such anyon systems is described by the Temperley-Lieb-Jones category TLJ($A$) for a fixed value of $A =\pm i \exp(\pm i\pi/{2(k+2)})$, and the associated topological quantum field theory is known as the Jones-Kauffman theory at level $k$ \cite{delaney2017}.\\
	
	The process of exchanging two adjacent anyons is known as braiding, which is a fundamental operation in anyon models. In particular, the braiding operation between the $n$-th and $(n+1)$th particles is referred to as the $\sigma_n$ braiding operator. The braid group is generated by the set of all possible braiding operators between adjacent anyons. These operators satisfy Artin relations, which include the Yang-Baxter equation or the type III Reidemeister move \cite{doi:10.1142/4256, delaney2017}. These relations ensure the algebraic consistency of anyon models \cite{Artin1947, Bakalov2000}. In addition to the hexagon and pentagon equations, Artin relations are practically useful in the implementation of unit-testing \footnote{Unit-testing involves evaluating each unit of the program separately to identify potential errors that could propagate to other parts of the program at an early stage \cite{Huizinga2007}.} for numerical packages used to compute braid generators. The verification process will be explained further in the next sections.
	The explicit matrix representation of a braid generator can be obtained by choosing a fusion space basis according to the fusion process and then applying the relevant $F$ and $R$ transformations \cite{Hormozi2007}.
	In general, there are two schemes to encode quantum information in the fusion states of anyons. In the dense encoding scheme, a minimal number of anyons is utilized. This scheme, which is discussed in Ref. \cite{PhysRevA.84.012332}, enables the systematic construction of two-qubit and three-qubit controlled phase gates. In the sparse encoding scheme, we represent each qubit or qudit with a definite number of anyons. Even though such construction does not allow for optimal use of the fusion space, it preserves the circuit model of quantum computation. {In this study, we develop a method for the sparse encoding scheme, and we give an algorithm to address the generalized case where anyons prepared in an arbitrary fusion state.} 
	In the following section, we will explicitly demonstrate how to compute the matrix representations of braid generators acting on the fusion state of identical anyons grouped in sets of a fixed number of anyons. Each set in our system consists of four anyons, since it is not recommended to represent a qubit with more than four anyons because of leakage, as proven for all anyon models in Ref. \cite{Ainsworth2011}. Nevertheless, the method {will be} extended systematically to an arbitrary number of anyons per set {in Appendix \ref{app:supplementary-materials}}. It is important to note that the fusion space of four anyons does not necessarily have the dimension of a qubit. Instead, it generally corresponds to a qudit.
	
	\section{Computation of Braid Generators for Sparse Encoding}
	
	\label{sec:calculating-breading-generators}
	
	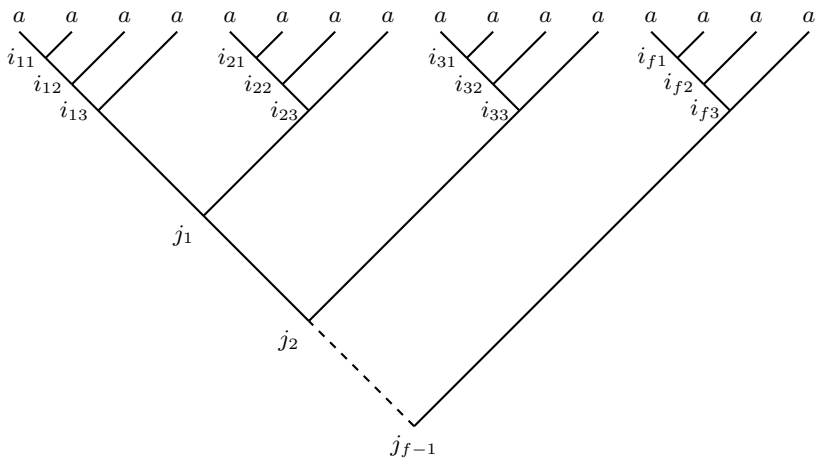
\begin{figure*}[t]
		\begin{center}
			
			\def\unitx{1.66}
			\def\unity{1.66}
			\def\minsize{9}
			\def\nq{4} 
			\def\na{4} 
			\newcommand{\fsize}{\Large}
			
			\resizebox{0.75\linewidth}{!}{
				\begin{tikzpicture}[
					roundnode/.style={circle, draw=blue!60, fill=blue!5, minimum size=\minsize mm},
					anyonnode/.style={circle, draw=red!60, fill=red!5, thick, minimum size=0.62*\minsize mm},
					]
					\def\orgnx{0};
					\def\orgny{0};
					
					\pgfmathsetmacro\nr{\na-1};
					\pgfmathsetmacro\nrq{\nq-1};
					
					\foreach \q in {0,...,\nrq}{
						\foreach \a in {1,...,\nr}{
							\def\rootx{\orgnx + \q * \unitx * \na + \a * \unitx/2};
							\def\rooty{\orgny -\unity/2 * \a};
							\draw[ultra thick]  (\rootx, \rooty) -- 
							(\rootx - \unitx/2, \rooty + \unity/2) ;
							\draw[ultra thick] (\rootx, \rooty)
							-- (\rootx + \a * \unitx/2, \orgny);
						};
					};
					
					\pgfmathsetmacro\orgnxr{\orgnx + \nr * \unitx/2};
					\pgfmathsetmacro\orgnyr{\orgny -\unity/2 * \nr};
					\pgfmathsetmacro\nrr{\nq-1};
					\pgfmathsetmacro\unitxr{\unitx * \na};
					\pgfmathsetmacro\unityr{\unity * \na};
					
					\foreach \a in {1,...,\nrr}{
						\def\rootx{\orgnxr + \a * \unitxr/2};
						\def\rooty{\orgnyr -\unityr/2 * \a};
						\ifthenelse{\a=\nrr}{
							\draw[ultra thick, loosely dotted, blue!60]  (\rootx, \rooty)
							-- (\rootx - \unitxr/2, \rooty + \unityr/2);
						}{
							\draw[ultra thick]  (\rootx, \rooty)
							-- (\rootx - \unitxr/2, \rooty + \unityr/2);
						}
						\draw[ultra thick] (\rootx, \rooty)
						-- (\rootx + \a * \unitxr/2, \orgnyr);
					};
					
					\foreach \q in {0,...,\nrq}{
						\foreach \a in {0,...,\nr}{
							\def\x{\orgnx + \q * \unitx * \na + \a * \unitx};
							\def\y{\orgny};
							\filldraw[black] (\x, \y) circle (0pt) node[anyonnode]{\fsize $a$};
						};
					};
					
					\foreach \q in {1,...,\nq}{
						\foreach \a in {1,...,\nr}{
							\pgfmathsetmacro\p{\q - 1};
							\def\rootx{\orgnx + \p * \unitx * \na + \a * \unitx/2};
							\def\rooty{\orgny -\unity/2 * \a};
							\ifthenelse{\q=\nq}{
								\filldraw[black] (\rootx, \rooty) circle (0pt) node[roundnode]{\fsize $i_{f\a}$};
							}{
								\filldraw[black] (\rootx, \rooty) circle (0pt) node[roundnode]{\fsize $i_{\q \a}$};
							}
						};
					};
					
					\foreach \a in {1,...,\nrr}{
						\def\rootx{\orgnxr + \a * \unitxr/2};
						\def\rooty{\orgnyr -\unityr/2 * \a};
						\ifthenelse{\a=\nrr}{
							\filldraw[black] (\rootx, \rooty) circle (0pt) node[roundnode]{\fsize $j_{f-1}$};
						}{
							\filldraw[black] (\rootx, \rooty) circle (0pt) node[roundnode]{\fsize $j_{\a}$};
						}
					};
				\end{tikzpicture}
			} 
		\end{center}
		\caption{This fusion tree diagram, depicted using the quantum circuit model of qubits, represents the state of $f$ groups, each containing four identical anyons and assigned a qudit. Here, ${i_{qp}}$ denotes the consecutive fusion outcomes of anyons inside the $q$-th qudit, where $q = 1, \ldots, f$ and $p = 1, \ldots, 3$. Additionally, $j_q$, where $q = 0, \ldots, (f-1)$, represents the fusion outcome of $i_{q3}$ and $i_{(q+1)3}$. It's important to note that $j_0 = i_{13}$ since, for negative $q$ values, $j_q$ represents the vacuum charge.}
		\label{fig:multi-qubit-state}
	\end{figure*}
	This {section is focused} on examining the matrix representation of braid generators that operate on a state $\ket{\psi}$ comprised of identical anyons arranged in groups of {multiple} anyons, with each group representing a single qubit/qudit. The diagrammatic representation of a typical state of this kind is depicted in Fig. \ref{fig:multi-qubit-state}. For simplicity, we can do all calculations for three qudits represented by four anyons each. Symbolically, such a state can be written in the full form as in Eq. \eqref{eq:full-form}. However, recognizing that this notation can be cumbersome, we express the previous identity in a more concise and simplified form as shown in Eq. \eqref{eq:concise-form}.
	\begin{table*}[t]
		\centering
		\begin{minipage}{\textwidth}
			\begin{align}
				\label{eq:full-form}
				\ket{\psi} &= \left| ((
				((((a,a)_{i_{11}}, a)_{i_{12}}, a)_{i_{13}}),  
				((((a,a)_{i_{21}}, a)_{i_{22}}, a)_{i_{23}}))_{j_1},
				((((a,a)_{i_{31}}, a)_{i_{32}}, a)_{i_{33}}))_{j_{2}}
				\right\rangle,\\
				\label{eq:concise-form}
				&= \left(
				\left(\ket{i_{11}, i_{12}, i_{13}} \otimes 
				\ket{i_{21}, i_{22}, i_{23}}\right)_{j_1} \otimes 
				\ket{i_{31}, i_{32}, i_{33}}
				\right)_{j_{2}}.
			\end{align}
		\end{minipage}
	\end{table*}
	In this identity, $\ket{i_{q1}, i_{q2}, i_{q3}}$ is a reduced notation for the state $\ket{(((a,a)_{i_{q1}}, a)_{i_{q2}}, a)_{i_{q3}}}$, which represents the state of the $q$-th set consisting of four anyons of type $a$. The symbols ${i_{q_1}, i_{q_2},i_{q_3}}$ denote the successive fusion results of the anyons within the $q$-th set.
	Additionally, the notation $\left(\ket{i_{11}, i_{12}, i_{13}} \otimes 
	\ket{i_{21}, i_{22}, i_{23}}\right)_{j_1}$ is a more convenient way of writing $\ket{i_{11}, i_{12}, i_{13}} \otimes 
	\ket{i_{21}, i_{22}, i_{23}} \otimes \ket{i_{13}, i_{23}, {j_1}}$ as defined in the Eq. \eqref{eq:state-tensor}. 
	The symbols ${j_1, j_2, \cdots, j_{f-1}}$ represent the fusion tree of the outcomes of the anyons ${i_{q3}}$, as shown in Fig. \ref{fig:multi-qubit-state}, where $q=1,2,\cdots,f$, and $f$ is the number of qudits, or the number of anyon groups present in the state $\ket{\psi}$. Also, $i_{q3}$ refers to the overall fusion outcome of the $q$-th anyon group.\\
	
	The matrix representation of a braid generator $\sigma_n$ is established by its constituent components, which are the likelihood amplitudes of converting any fusion eigenstate $\ket{i}$ to another fusion eigenstate $\ket{j}$ within a specified basis by applying the specified operation. This is expressed as follows:
	\begin{align}
		\label{eq:braiding_matrix_components}
		\left[\sigma_n\right]_{ji} &= \bra{j}\sigma_n \ket{i}.
	\end{align}
	Ultimately, applying $\sigma_n$ to a specific fusion eigenstate falls into two cases. In the first case, $\sigma_n$ interchanges two adjacent anyons within the same qudit group. The amplitude for this process can be calculated using the standard braiding matrix $B$ \cite{Pachos2012, Field2018}. In the second case, $\sigma_n$ exchanges two adjacent anyons belonging to different qudit groups. For this scenario, we introduce a knitting matrix $K$. The detailed calculations for each case will be demonstrated in the following subsections.
	
	\subsection{The braiding matrix $B$}
	\label{sec:braiding-matrix-B}
	Consider the case of interchanging two adjacent anyons within the same qudit set i.e the braid generator index $n$ is not a multiple of $4$. Hence, the braiding operation is only applied to the $q$-th qudit, where $q=n|4+1$ and $n|4$ is the {Euclidean quotient of $n$ divided by $4$}. Consequently, applying $\sigma_n$ to the state $\ket{\psi}$ is equivalent to implementing $\sigma_{(n \text{ mod } 4)}$ on the state of the $q$-th anyon group, where $n \text{ mod } 4$ is the residue {of $n$ divided by $4$}. Let us say that $q = 2$. Then,
	\begin{align}
		\sigma_n \ket{\psi} 
		=
		\Big( 
		\big(\ket{i_{11}, i_{12}, i_{13}}
		\otimes
		\sigma_{(n \text{ mod } 4)} \ket{i_{21}, i_{22}, i_{23}}
		\big)_{j_1}
		\otimes 
		\ket{i_{31}, i_{32}, i_{33}}
		\Big)_{j_{2}}.
	\end{align}
	In the case where $n$ satisfies the congruence $n = 1 \pmod{4}$, the $R$ matrix definition can be used to determine the action of the braiding operator $\sigma_{1}$ on a basis state of four anyons, denoted as $\ket{i,j,k}$. Specifically, the action of $\sigma_{1}$ on $\ket{i,j,k}$ can be expressed using the tensorial form \eqref{eq:state-tensor} and the definition of the $R$ matrix given in Eq. \eqref{eq:R_matrix_def} as
	\begin{align}
		\sigma_{1}\ket{i,j,k} 
		=& \sigma_1 \ket{(((a,b)_i,c)_j,d)_k}, \nonumber\\
		=& R_{ab}^{i} \ket{i,j,k}.
	\end{align}
	Therefore, the diagonal elements of the matrix $[\sigma_1]$ are  equal to $R^i_{ab}$.\\
	
	When $n$ satisfies the congruence $n = 2 \pmod{4}$, the action of $\sigma_2$ on a four-anyon basis state $\ket{i,j,k}$ can be determined using the $F$ and $R$ transformations. Specifically, the action of $\sigma_2$ on $\ket{i,j,k}$ can be expressed as
	\begin{align}
		\label{eq:sigma_2_00}
		\sigma_{2} \ket{i,j,k}
		=& \sigma_2 \ket{(((a,b)_i,c)_j,d)_k}, 
		\nonumber\\
		=& \sum_{ml} (F_{abc}^j)_l^i R_{bc}^{l} (F_{acb}^{\dagger j})_m^l \ket{(((a,c)_m,b)_j, d)_k},
	\end{align}
	which involves a linear product of matrices. Since the $R$ matrix is diagonal, Eq. \eqref{eq:sigma_2_00} can be written in terms of the braiding matrix elements as:
	\begin{align}
		\label{eq:s2}
		\sigma_2 \ket{(((a,b)_i,c)_j,d)_k} =
		\sum_m  \left( B^j_{abc} \right)_m^i \ket{(((a,c)_m,b)_j, d)_k},
	\end{align}
	where
	\begin{align}
		\label{eq:braiding-op}
		\left( B^j_{abc} \right)_m^i =
		\sum_l (F_{abc}^j)_l^i R_{bc}^{l} (F_{acb}^{\dagger j})_m^l \ ,
	\end{align}
	with the braiding matrix $B_{abc}^j$ being defined as $B^j_{abc} = F_{abc}^j R_{bc} F_{acb}^{\dagger j}$.
	Note that in contrast to $\sigma_1$, which is limited to phase operations and cannot perform superpositions, the braiding operator $\sigma_2$ can mix between the fusion eigenstates and execute the desired superposition operations in the chosen fusion basis.\\
	
	In the case where $n = 3 \pmod{4}$, the braiding operator $\sigma_3$ can be obtained by applying the operator $\sigma_2$ to the state $\ket{((i,c)_j, d)_k}$. This means that, based on the expression in Eq. \eqref{eq:s2}, we have:
	
	\begin{align}
		\sigma_3 \ket{i,j,k} 
		&= \sum_m \left( B^k_{icd} \right)_m^j \ket{(((a,b)_i,d)_m,c)_k}.
	\end{align}
	In general, the braiding matrix $B_{abc}^i$ computes all possible braiding operations between two adjacent anyons within the same qudit group, even if the number of anyons per qudit is greater than four. Namely,
	\begin{align}
		\label{eq:s2}
		\sigma_2 \ket{(((a,b)_i,c)_j,d)_k} =
		\sum_m  \left( B^j_{abc} \right)_m^i \ket{(((a,c)_m,b)_j, d)_k},
	\end{align}
	Furthermore, $\sigma_1$ can be represented as $B_{\vac ab}^a$.
	\subsection{The knitting matrix $K$}
	\label{eq:mixing-matrix-M}
	In the second case, we assume that $n$ is a multiple of four ($n=4m$). In this scenario, the operator $\sigma_n$ acts on the joint state of the $m$-th and $(m+1)$th neighboring groups by exchanging the adjacent edge anyons. To apply this operator, one may first use an $F$ transformation to decouple the joint state of the two groups. For simplicity, let us illustrate the steps for $m = 2$ in which case the braiding operator writes Eq. \eqref{eq:sigma_n_multi-qudits} and \eqref{eq:sigma_n_multi-qudits-F}.
	\begin{align}
		\label{eq:sigma_n_multi-qudits}
		\sigma_{n} \ket{\psi}
		=& \sigma_{4 \times 2} \left(
		\left(\ket{i_{11}, i_{12}, i_{13}} \otimes 
		\ket{i_{21}, i_{22}, i_{23}}\right)_{j_1} \otimes 
		\ket{i_{31}, i_{32}, i_{33}}
		\right)_{j_{2}},\\
		\label{eq:sigma_n_multi-qudits-F}
		=&  \sum_k \left(F_{j_{0} i_{23} i_{33}}^{j_2}\right)^{j_{1}}_k
		\left(
		\ket{i_{11}, i_{12}, i_{13}} \otimes 
		\sigma_4
		\left(
		\ket{i_{21}, i_{22}, i_{23}} \otimes 
		\ket{i_{31}, i_{32}, i_{33}}
		\right)_k
		\right)_{j_{2}}.
	\end{align}
	%
	Notice that in Eq. \eqref{eq:sigma_n_multi-qudits-F}, $j_0 = i_{13}$ and in general $j_n$ represents the vacuum charge for all $n<0$. In doing so, the scenario has shifted to the application of the braiding operation on a shared state of two neighboring qudits. By repeatedly performing $F$ moves, the resulting sequence of transformations on the joint state of these two qudits will be as follows:
	\begin{align}
		\left(
		\ket{i_{21}, i_{22}, i_{23}} \otimes 
		\ket{i_{31}, i_{32}, i_{33}}
		\right)_k 
		&=
		\ket{((((a,a)_{i_{21}},a)_{i_{22}}, a)_{i_{23}}, (((a,a)_{i_{31}},a)_{ i_{32}}, a)_{i_{33}})_k}, \nonumber\\
		&=
		\sum_{p_3 p_{2} p_1}
		\left(F_{i_{23}i_{32}a}^{\dagger k}\right)^{i_{33}}_{p_3}
		\left(F_{i_{23}i_{31}a}^{\dagger p_3}\right)^{i_{32}}_{p_{2}}
		\left(F_{i_{23}aa}^{\dagger p_2}\right)^{i_{31}}_{p_1} \nonumber\\
		&\times \ket{(((((((a,a)_{i_{21}},a)_{ i_{22}}, a)_{i_{23}}, a)_{p_{1}},a)_{p_{2}}, a)_{p_3}, a)_k}.
	\end{align}
	Upon application of the braiding matrix $B$ as defined in Eq. \eqref{eq:braiding-op} to the state {$\ket{((i_{m2},a)_{i_{m3}},a)_{p_1}}$},  an immediate outcome can be obtained, which is as follows:
	\begin{align}
		\sigma_4 (
		\ket{i_{21}, i_{22}, i_{23}} \otimes 
		\ket{i_{31}, i_{32}, i_{33}}
		)_k
		&= \sum_{p_3 p_{2} p_1 i'_{23}}
		\prod_{r=1}^{3} \left(F_{i_{23}i_{3,3-r}a}^{\dagger p_{3-r+2}} \right)^{i_{3,4-r}}_{p_{4-l}}
		\sum_{i'_{23}}
		\left(B_{i_{22}aa}^{p_1}\right)^{i_{23}}_{i'_{23}} \nonumber\\ &\times
		\ket{(((((((a,a)_{i_{21}},a)_{ i_{22}}, a)_{i'_{23}}, a)_{p_1},a)_{p_2}, a)_{p_3}, a)_k}.
	\end{align}
	such that $p_4 = k$. The subsequent step involves the conversion of the state back into its original basis form, which can be achieved by employing the suitable $F$ transformations and the braiding matrix $B$, as shown in Eq. \eqref{eq:L}. Thus, the resulting transformation is {represented} by a new braiding matrix $L$, which represents the effect of braiding two edge anyons in a joint state of neighboring qudits. The $L$ matrix is defined in Eq. \eqref{eq:L-definition}.
	
	\begin{align}
		\label{eq:L}
		\sigma_4 (
		\ket{i_{21}, i_{22}, i_{23}} &\otimes 
		\ket{i_{31}, i_{32}, i_{33}}
		)_k = \nonumber \\
		&\sum_{i'_{23}i'_{31}i'_{32}i'_{33}}
		\left(L^k_{i_{22}}\right)^{i_{23} i_{31} i_{32} i_{33}}_{i'_{23} i'_{31}  i'_{32} i'_{33}}
		(
		\ket{i_{21}, i_{22}, i'_{23}} \otimes 
		\ket{i'_{31}, i'_{32}, i'_{33}}
		)_k,
	\end{align}
	\begin{align}
		\label{eq:L-definition}
		\left(L^{k}_{i_{22}}\right)^{i_{23} i_{31} i_{32} i_{33}}_{i'_{23} i'_{31}  i'_{32} i'_{33}}
		= \sum_{p_3 p_{2} p_1 i'_{23}} 
		\prod_{r=1}^{3} \left(F_{i_{23}i_{3,3-r}a}^{\dagger p_{3-r+2}} \right)^{i_{3,4-r}}_{p_{4-r}}
		\left(B_{i_{22}aa}^{p_1}\right)^{i_{23}}_{i'_{23}} 
		\prod_{r=1}^{3} \left(F_{i'_{23}i_{3,3-r}a}^{p_{3-r+2}} \right)_{i'_{3,4-r}}^{p_{4-r}}.
	\end{align}
	
	After performing computations on the joint state of the two qudits, it is necessary to transform the resulting state back to the original form specified by Eq. \eqref{eq:sigma_n_multi-qudits} using the inverse $F$ transformations. Additionally, it is necessary to introduce another braiding matrix, denoted by $K$ the knitting matrix, that accounts for all feasible fusion states achieved by exchanging anyons shared by adjacent qudit groups. This is demonstrated in Eq. \eqref{eq:knitting}. The components of the introduced matrix $K$ are defined by a specific linear combination of the components of $L$ as shown in Eq. \eqref{eq:knitting-definition}.
	
	\begin{align}
		\label{eq:knitting}
		&\sigma_{4\times 2} \ket{\psi} = \nonumber \\
		&  \sum_{j'_{1} i'_{23}i'_{31}i'_{32} i'_{33}} 
		\left(K^{j_0 j_2}_{i_{22}}\right)^{j_1 i_{23} i_{31} i_{32} i_{33}}_{j'_{1} i'_{23}i'_{31}i'_{32} i'_{33}}
		\left(
		\left(\ket{i_{11}, i_{12}, i_{13}} \otimes 
		\ket{i_{21}, i_{22}, {i'}_{23}}\right)_{{j'}_1} \otimes 
		\ket{{i'}_{31}, {i'}_{32}, {i'}_{33}}
		\right)_{j_{2}}.
	\end{align}
	\begin{align}
		\label{eq:knitting-definition}
		&\left(K^{j_0 j_2}_{i_{22}}\right)^{j_1 i_{23} i_{31} i_{32} i_{33}}_{j'_{1} i'_{23}i'_{31}i'_{32} i'_{33}} =
		\sum_k
		\left(F_{j_{0} i_{23} i_{33}}^{j_2}\right)^{j_{1}}_k 
		\left(L^k_{i_{22}}\right)^{i_{23} i_{31} i_{32} i_{33}}_{i'_{23} i'_{31}  i'_{32} i'_{33}}
		\left(F_{j_{0} i'_{23} i'_{33}}^{\dagger j_2}\right)_{j'_{1}}^k.
	\end{align}
	
	In summary, by computing the $B$, $L$, and $K$ matrices from the $F$ and $R$ matrices, we can determine the effect of any braiding operator acting on a state of multiple groups of anyons, where each group represents a qudit. In other words, the calculation of the right-hand side of Eq. \eqref{eq:braiding_matrix_components} has become straightforward for any number of qubits due to the availability of the aforementioned transformations. The general formula for the $L$ and $K$ matrices, along with their application to a fusion state involving an arbitrary number of anyons per qudit and arbitrary size, are provided in Appendix. \ref{app:supplementary-materials}.
	
	In order to ensure the reliability of our method, it is crucial to meet rigorous testing conditions that guarantee the accuracy of the formulas and their programmatic implementation.  To achieve this, we will refer to the algebra of the braid group known as Artin relations, which serves as a foundation for our approach \cite{Artin1947}:
	\begin{equation}
		\begin{array}{rccc}
			\sigma_{i} \sigma_{i+1} \sigma_{i} & = & \sigma_{i+1} \sigma_{i} \sigma_{i+1} \ , & \\
			\sigma_{i} \sigma_{j} & = & \sigma_{j} \sigma_{i} \ , & \left|i-j\right| > 1,\\
			\sigma_i \sigma_i^{-1} & = & I. &
		\end{array}
	\end{equation}
	To verify Artin relations numerically, we should ensure that the distance between the left and right sides of the relations is negligible, meaning it is in the order of the floating-point precision of the specific device and programming language being used. Additionally, it is necessary to substitute the third relation with the unitarity condition since our focus is on quantum computing applications \cite{delaney2017}. In order to calculate the difference between two unitaries, we employ the spectral distance, which will be defined in the upcoming section. We have performed numerical verification to validate the formulas presented in this section using Artin algebra for up to 24 anyons within the Fibonacci and Ising anyon models. The tests reveal a difference in the order of {$10^{-15}$} between the left and right sides of the relations.
	In addition, it is convenient to {use this method to simulate numerically two exemplary cases: a CNOT gate using six Fibonacci anyons and a GHZ state using fifteen Fibonacci anyons. This will be the main goal of sections \ref{sec:cnot} and \ref{sec:ghz}.}

	\section{Computation of Braid Generators: A General Method}
	Although the knitting matrix is helpful for computing the braid generator matrices when the fusion states are formatted in a specific order as illustrated in Fig. \ref{fig:multi-qubit-state}, it may be necessary in general to compute the braid generators systematically for any fusion order. In this scenario, we will conduct the computation in two steps: First, we transform the state into the standard basis form, which is characterized by the successive fusion process ordered from left to right. Second, we utilize the standard braiding matrix $B$ to compute the braid generator, as explained in Section \ref{sec:braiding-matrix-B}. The consisting challenge is how to make the first step systematic, implementable by a machine and optimal.\\
	Initially, note that an arbitrary fusion process can be represented by a binary tree within the context of graph theory. In this representation, the fusion process forms a graph where nodes denote anyon charges connected by lines. Since loops and high fusion multiplicities do not matter, the structure forms a tree. Furthermore, the fusion tree features a root node representing the total anyon charge, establishing it as a rooted tree. Additionally, since the order of fused anyons matters, it qualifies as an ordered tree. 
	Finally, each elementary fusion process involves only two branches and one parent, thus forming a binary tree. Conventionally, navigation through the binary tree is facilitated by specifying left and right directions from each step starting at the root.
	
	Interestingly, assuming that every node retains information about its parent and its two branches, each node in the fusion tree stores recursive information about the entire tree. Consequently, useful definitions can be established. We define a node as an entity possessing an anyonic charge, with at most one parent and at most two branches. The root is the node without a parent, while the leaves are nodes lacking children. We define composite nodes as those that are not leaves. If a node is labeled as $N$, we define the functions pointing to the left and right branches as $\text{L}(N)$ and $\text{R}(N)$, respectively. The depth of a node $\text{D}(N)$ is its distance from the root. Furthermore, we introduce the functional power $\text{Power}(f, m) = f^m$, which composes the function $f$ with itself $m$ times, denoted as $f^m = \underbrace{f \circ \cdots \circ f}_{m}$. For instance, $L^2(\text{Root})=\text{L}(\text{L}(\text{Root}))$, representing the left-most node $M$ with depth $D(M)=2$. For practical reasons, amplitudes can be attributed to the trees, generally involving $F$ components contracted with nodes.
	
	It is worth noting that the optimal number of moves required to transform an arbitrary fusion state into the standard form cannot be less than the number of unwanted branches. These unwanted branches comprise all left branches not conforming to the form of $L^i(\text{Root})$ for $i$ be a positive integer. Here, we present an algorithm that transforms any fusion tree into the standard form, specifying all required $F$ amplitudes, with a number of moves equal to the number of unwanted branches, while the number of steps is proportional to the number of composite nodes in the tree. This algorithm is based on the routine algorithm described in Algorithm \ref{alg:right-transformation}, which standardizes the right branch of any given fusion state by simply applying recursive $F^\dagger$ moves on the right branch until it becomes a leaf node. Consequently, the main algorithm is based on iteratively applying the routine Algorithm \ref{alg:right-transformation} on the left-most branches. The step-by-step procedure to implement this basic idea is detailed in Algorithm \ref{alg:to-standard}. 
	To better understand the procedure, it is helpful to examine two special cases. The first one involves entering a fusion state in the standard form. In this scenario, the routine algorithm will not affect any node because all nodes' right branches are leaves. The second case involves considering the symmetrically reversed standard form where all nodes' left branches are leaves. Initially, the algorithm will apply the routine algorithm to the root node, which recursively transforms the fusion tree into the standard form. After that, iterating over the created left nodes will be identical to the first case with no additional effects.
	\begin{algorithm}
		\caption{StandardizeRightBranch: This is a recursive algorithm that aims to transform the right branch of the input fusion state into a leaf. This is achieved by recursively transforming the successor right branches into leaves until the condition is met. Meanwhile, the CoefficientsList is populated with the required $F^\dagger$ amplitudes to be retained.}
		\label{alg:right-transformation}
		\begin{algorithmic}[1]
			\Require CoefficientsList, $\ket{(a, b)_c}$ such that $b$ is composite.
			\Ensure CoefficientsList, $\ket{(a, b)_c}$ such that $b$ is not composite.
			\If{$b$ is \textit{not} composite}
			\State \Return $\ket{(a, b)_c}$
			\Else
			\State let $\ket{(a, b)_c} = \ket{(a, (e, f)_b)_c}$ \Comment{Since $b$ is compsite.}
			\State $l \gets \text{Length(CoefficientsList)}$
			\State $\text{CoefficientsList} \gets \text{CoefficientsList} + (F_{aef}^{\dagger c})^b_{b_l}$
			\State $\ket{(a, b)_c} \gets \ket{((a, e)_{b_l}, f)_c}$ \Comment{Update the state.}
			\EndIf
			\While{$b$ is composite}
			\State CoefficientsList, $\ket{(a, b)_c} \gets \text{StandardizeRightBranch}(\text{CoefficientsList}, \ket{(a, b)_c})$
			\EndWhile
			\State \Return CoefficientsList, $\ket{(a, b)_c}$
		\end{algorithmic}
	\end{algorithm}
	
	\begin{algorithm}
		\caption{StandardizeFusionState: The purpose of this algorithm is to make the transformation from an arbitrary fusion state to the standard fusion state format specifying all necessary $F$ amplitudes. The basic idea is to apply the routine Algorithm \ref{alg:right-transformation} on all the left most nodes and updating the fusion state in each iteration. $L^d(c)$ is a function that points to the left-most node of distance $d$ from the root $c$.}
		\label{alg:to-standard}
		\begin{algorithmic}[1]
			\Require CoefficientsList, $\ket{(a, b)_c}$.
			\Ensure CoefficientsList, $\ket{(a, b)_c}$ in standard form.
			\State $\text{CoefficientsList},\ket{(a, b)_c} \gets \text{StandardizeRightBranch}(\text{CoefficientsList}, \ket{(a, b)_c})$
			\State $i \gets 0$
			\State $g \gets L^i(c))$ \Comment{At this step, $g=c$}
			\State let $\ket{g} = \ket{(e, f)_g}$ \Comment{At this step, $e=a$ and $f=b$}
			\While{$e$ is composite}
			\State let $\ket{e} = \ket{(k, l)_e}$ \Comment{$\ket{g} = \ket{((k, l)_e, f)_g}$}
			\State $\text{CoefficientsList}, \ket{(k, l)_e} \gets \text{StandardizeRightBranch}(\text{CoefficientsList}, \ket{(k, l)_e})$
			\State $L^i(c) \gets g$ \Comment{Update the fusion tree.}
			\State $i \gets i+1$
			\State $g \gets L^i(c)$
			\State let $\ket{g} = \ket{(e, f)_g}$
			\EndWhile
		\end{algorithmic}
	\end{algorithm}
	
	\section{Implementation of Topological CNOT Gate}
	\label{sec:cnot}
	
	In this section, our primary objective is to {compute} the braid of the CNOT gate, initially introduced by Bonesteel et al., utilizing the injection method \cite{Bonesteel2005}, {as a validation of our approach}. {We specifically chose this braid because it has been proven to be accurate whenever the single-qubit gates being substituted, namely the injection and target gates, are well approximated. Moreover, the CNOT gate will be used in the next section for GHZ state compilation.} 
	
	\begin{table}[t]
		\caption{There are exclusively three allowed fusion states for three Fibonacci anyons. We can assign a logical state to each one of them. However, the states $\ket{0}$ and $\ket{1}$ are linearly independent of $\ket{2}$ since their sectors, i.e., global fusion outcomes, are different. Consequently, in the sector $\fib$, we can {precisely} represent a qubit, without {introducing additional} dimensions.
			However, the Hilbert space of three Fibonacci anyons in the sector $\vac$ is trivial and cannot represent a qubit.}
		\label{tab:basis-1q}
		\begin{tabular*}{\linewidth}{@{\extracolsep\fill}ccr}
			\toprule
			Index & Logical State & Fusion State                           \\
			\midrule
			$0$   & $\ket{2}$     & $\ket{((\fib, \fib)_\fib, \fib)_\vac}$ \\
			$1$   & $\ket{0}$     & $\ket{((\fib, \fib)_\vac, \fib)_\fib}$ \\
			$2$   & $\ket{1}$     & $\ket{((\fib, \fib)_\fib, \fib)_\fib}$ \\
			\bottomrule
		\end{tabular*}
	\end{table}
	\paragraph*{The anyon model.} In \cite{Bonesteel2005}, the CNOT quantum gate is designed using six Fibonacci anyons with three anyons per qubit. The Fibonacci model includes one anyonic charge labeled as $\fib$ and the vacuum $\vac$, and the non-trivial fusion rule of these anyons is $\fib \times \fib = \vac + \fib$ \cite{Rouabah2021}. Hence, given three anyons, the possible fusion states are as shown in Tab. \ref{tab:basis-1q}.
	Note that the logical states $\ket{0}$ and $\ket{1}$ are naturally independent of $\ket{2}$ due to their different global fusion outcomes. Taking into consideration that a single qubit can be effectively represented by three Fibonacci anyons, resulting in a collective outcome of  $\fib$,  it follows that the state  $\ket{2}$  is non-computational.
	Considering six anyons, the fusion basis can be ordered as shown in Tab. \ref{tab:basis-2q}.
	Similarly, we have two separate sectors in the fusion space, which are differentiated by their global fusion outcomes. Consequently, we can choose one of two possible computational bases: $\left\lbrace \ket{00}_\vac \ket{10}_\vac \ket{01}_\vac \ket{11}_\vac \right\rbrace$ or $\left\lbrace \ket{00}_\fib \ket{10}_\fib \ket{01}_\fib \ket{11}_\fib \right\rbrace$, while all other states that include one of the qubits in state $\ket{2}$ are considered non-computational.\\ 
	\begin{table}[t]
		\caption{The fusion states presented in this table represent the exclusive possible states with six Fibonacci anyons. Logical states can be assigned to each fusion state, considering the sector of these states, i.e., the overall outcome. It is observed that a computational space of four states, namely $\left\lbrace \ket{00}_\vac \ket{10}_\vac \ket{01}_\vac \ket{11}_\vac \right\rbrace$ or $\left\lbrace \ket{00}_\fib \ket{10}_\fib \ket{01}_\fib \ket{11}_\fib \right\rbrace$ in the $\vac$ and $\fib$ sectors respectively, can represent two qubits. An index is assigned to each state to facilitate the labeling of braid matrices.}
		\label{tab:basis-2q}
		\begin{tabular*}{\linewidth}{@{\extracolsep\fill}ccc}
			\toprule
			Index & Logical State & Fusion State\\
			\midrule
			$0$   & $\ket{22}_\vac$   & $\ket{(((\fib, \fib)_{\fib}, \fib)_{\vac}, ((\fib, \fib)_{\fib}, \fib)_{\vac})_\vac}$\\
			$1$   & $\ket{00}_\vac$   & $\ket{(((\fib, \fib)_{\vac}, \fib)_{\fib}, ((\fib, \fib)_{\vac}, \fib)_{\fib})_\vac}$\\
			$2$   & $\ket{10}_\vac$   & $\ket{(((\fib, \fib)_{\fib}, \fib)_{\fib}, ((\fib, \fib)_{\vac}, \fib)_{\fib})_\vac}$\\
			$3$   & $\ket{01}_\vac$   & $\ket{(((\fib, \fib)_{\vac}, \fib)_{\fib}, ((\fib, \fib)_{\fib}, \fib)_{\fib})_\vac}$\\
			$4$   & $\ket{11}_\vac$   & $\ket{(((\fib, \fib)_{\fib}, \fib)_{\fib}, ((\fib, \fib)_{\fib}, \fib)_{\fib})_\vac}$\\
			$5$   & $\ket{02}_\fib$   & $\ket{(((\fib, \fib)_{\vac}, \fib)_{\fib}, ((\fib, \fib)_{\fib}, \fib)_{\vac})_\fib}$\\
			$6$   & $\ket{12}_\fib$   & $\ket{(((\fib, \fib)_{\fib}, \fib)_{\fib}, ((\fib, \fib)_{\fib}, \fib)_{\vac})_\fib}$\\
			$7$   & $\ket{20}_\fib$   & $\ket{(((\fib, \fib)_{\fib}, \fib)_{\vac}, ((\fib, \fib)_{\vac}, \fib)_{\fib})_\fib}$\\
			$8$   & $\ket{21}_\fib$   & $\ket{(((\fib, \fib)_{\fib}, \fib)_{\vac}, ((\fib, \fib)_{\fib}, \fib)_{\fib})_\fib}$\\
			$9$   & $\ket{00}_\fib$   & $\ket{(((\fib, \fib)_{\vac}, \fib)_{\fib}, ((\fib, \fib)_{\vac}, \fib)_{\fib})_\fib}$\\
			$10$  & $\ket{10}_\fib$   & $\ket{(((\fib, \fib)_{\fib}, \fib)_{\fib}, ((\fib, \fib)_{\vac}, \fib)_{\fib})_\fib}$\\
			$11$  & $\ket{01}_\fib$   & $\ket{(((\fib, \fib)_{\vac}, \fib)_{\fib}, ((\fib, \fib)_{\fib}, \fib)_{\fib})_\fib}$\\
			$12$  & $\ket{11}_\fib$   & $\ket{(((\fib, \fib)_{\fib}, \fib)_{\fib}, ((\fib, \fib)_{\fib}, \fib)_{\fib})_\fib}$\\
			\bottomrule
		\end{tabular*}
	\end{table}
	\newcommand{\picwidth}{0.19\linewidth}
	\begin{figure*}[t!]
		\centering
		\begin{subfigure}[b]{\picwidth}
			\centering
			\includegraphics[width=\textwidth]{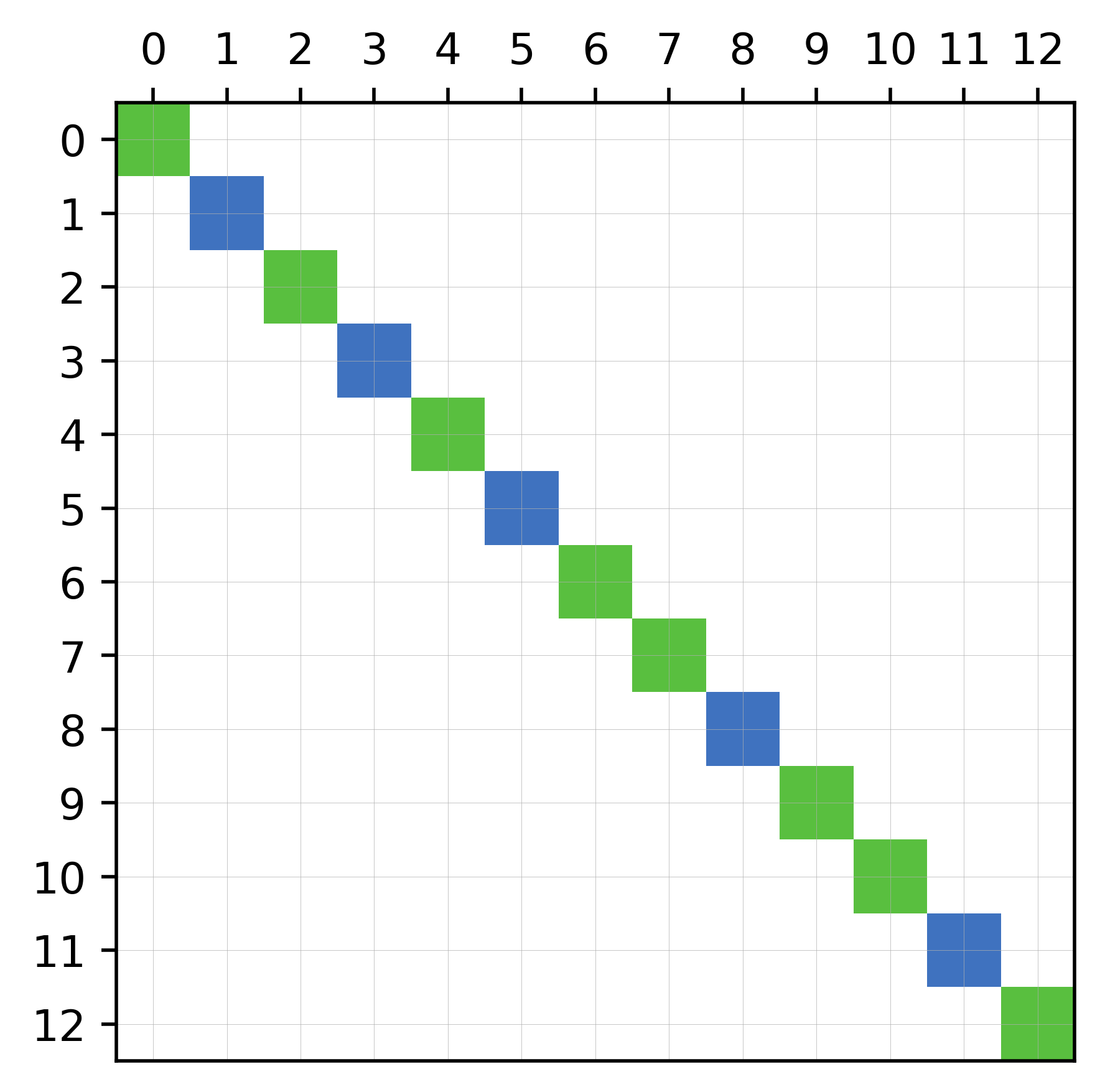}
			\caption*{$\sigma_1$}
		\end{subfigure}
		\begin{subfigure}[b]{\picwidth}
			\centering
			\includegraphics[width=\textwidth]{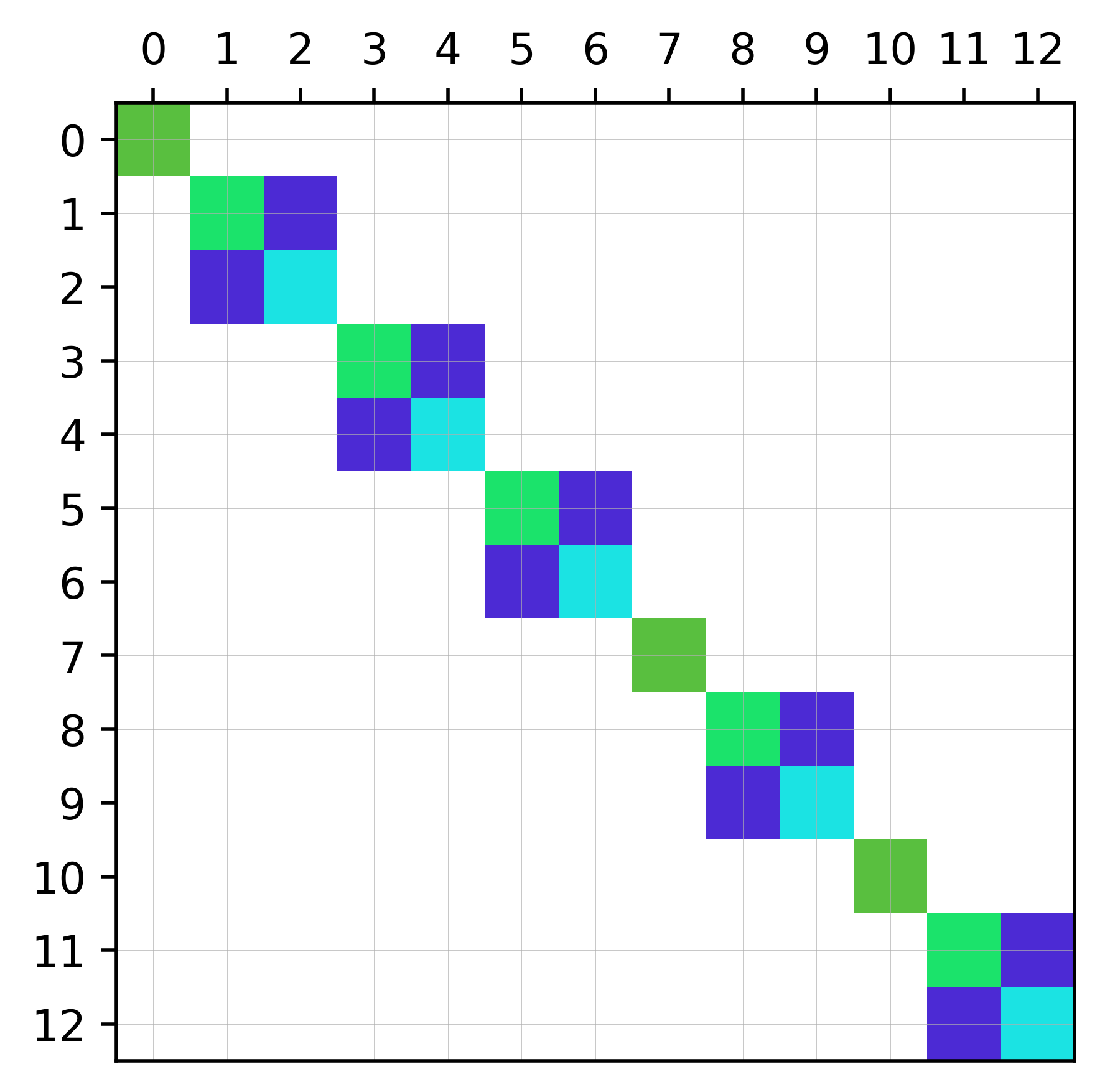}
			\caption*{$\sigma_2$}
		\end{subfigure}
		\begin{subfigure}[b]{\picwidth}
			\centering
			\includegraphics[width=\textwidth]{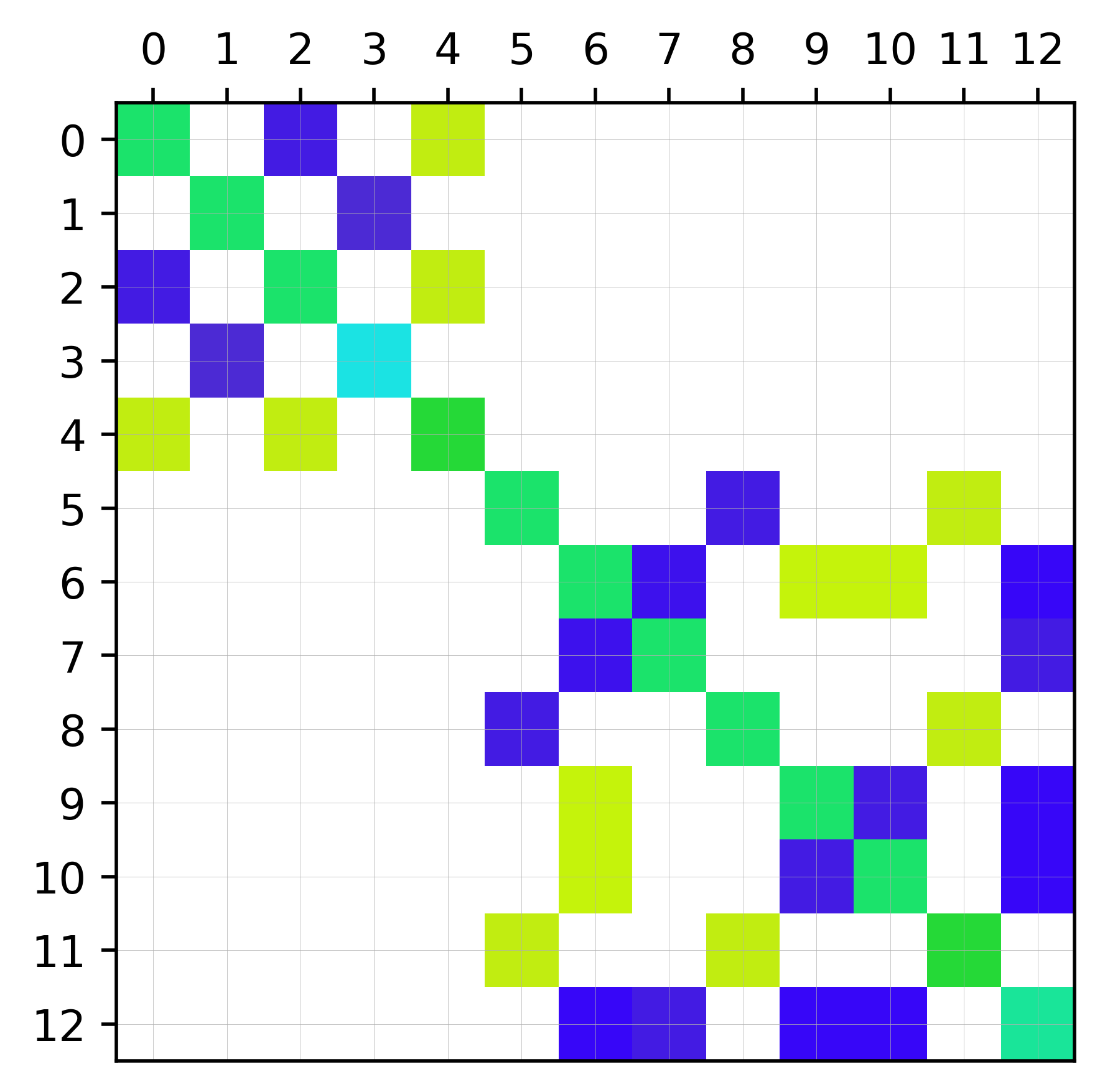}
			\caption*{$\sigma_3$}
		\end{subfigure}
		\begin{subfigure}[b]{\picwidth}
			\centering
			\includegraphics[width=\textwidth]{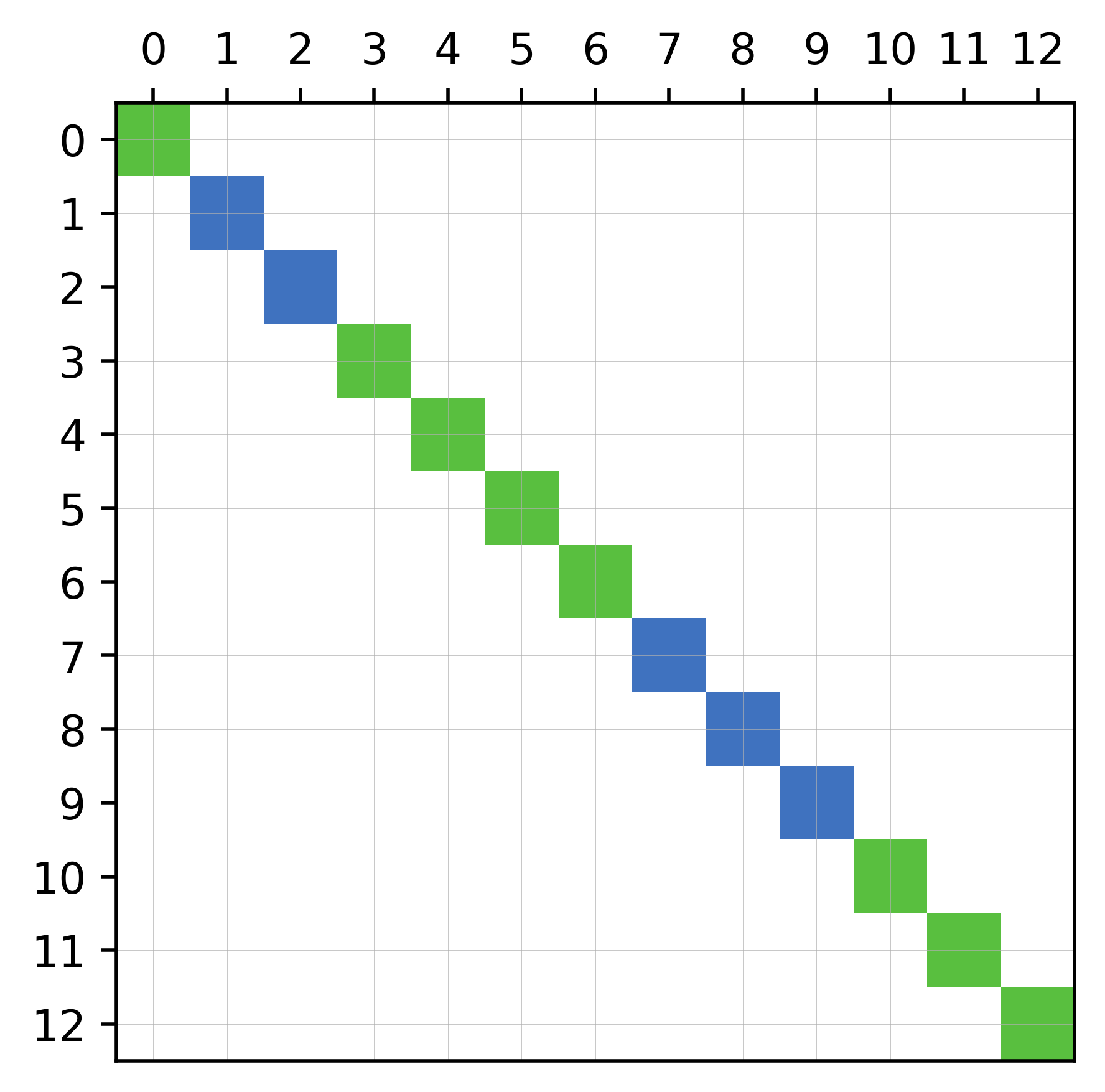}
			\caption*{$\sigma_4$}
			\label{fig:toffoli-controlled-injection}
		\end{subfigure}
		\begin{subfigure}[b]{\picwidth}
			\centering
			\includegraphics[width=\textwidth]{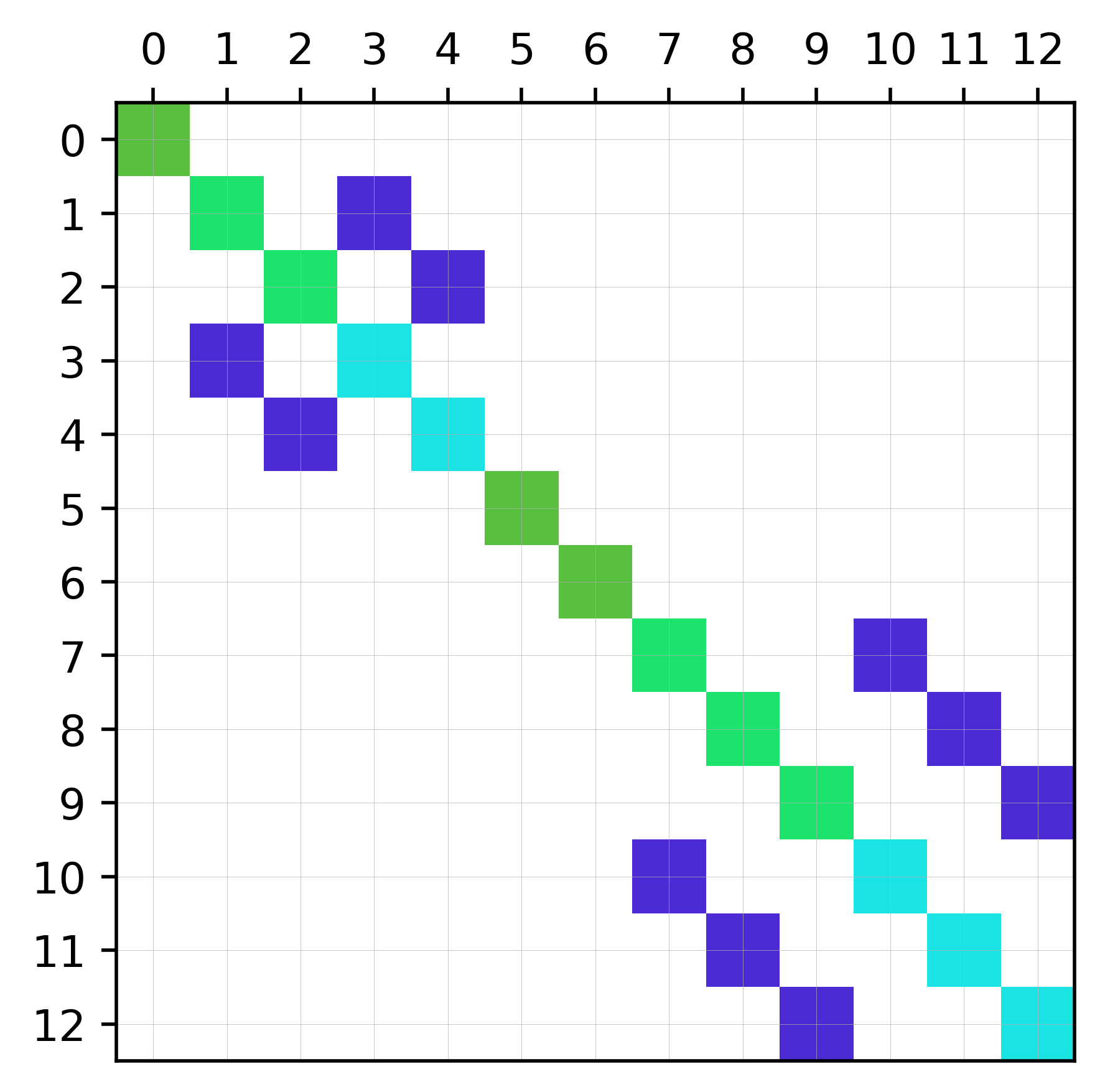}
			\caption*{$\sigma_5$}
			\label{fig:xor}
		\end{subfigure}
		\begin{subfigure}[b]{\picwidth}
			\centering
			\includegraphics[width=\textwidth]{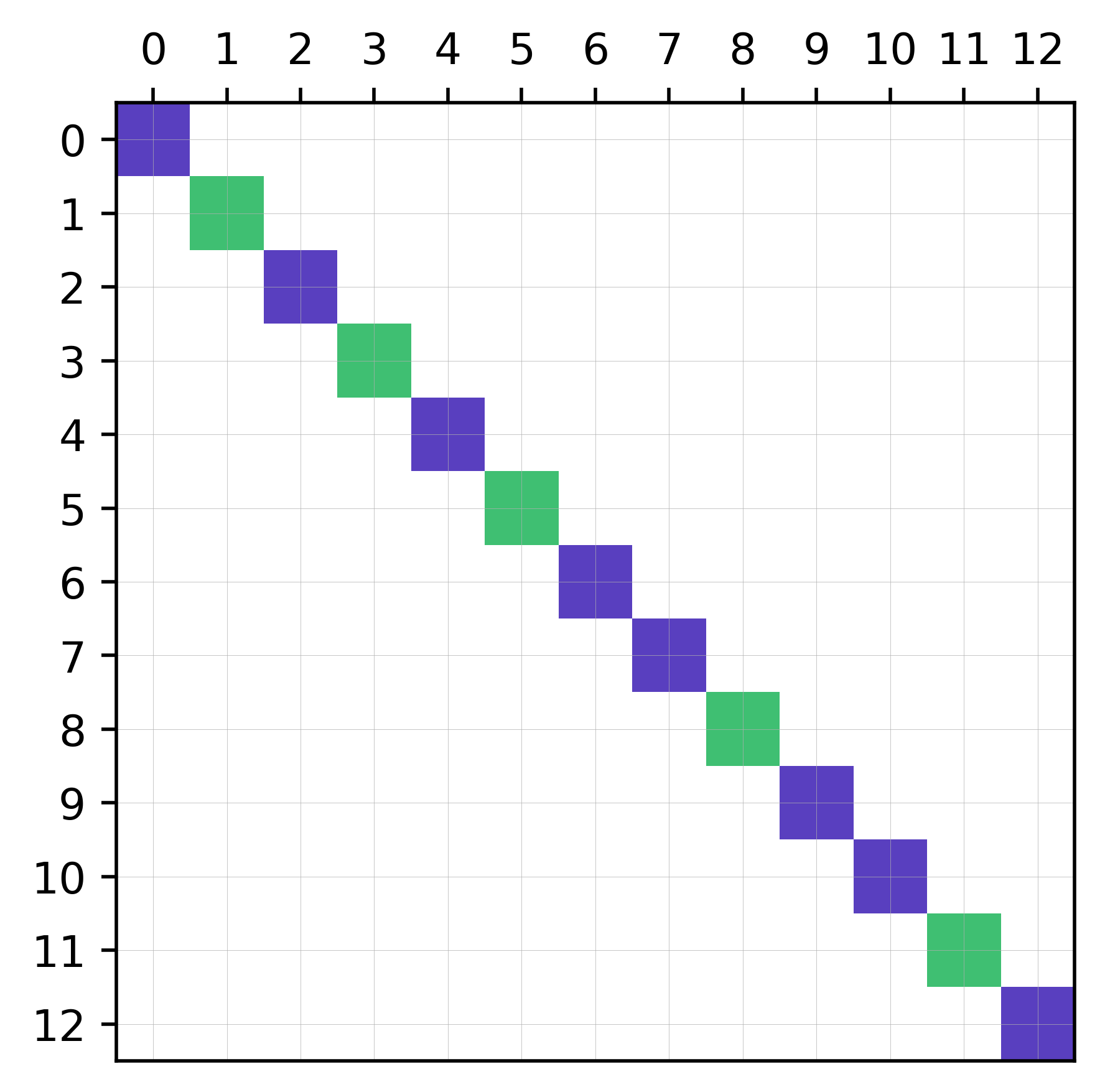}
			\caption*{$\sigma_1^{-1}$}
			\label{fig:and}
		\end{subfigure}
		\begin{subfigure}[b]{\picwidth}
			\centering
			\includegraphics[width=\textwidth]{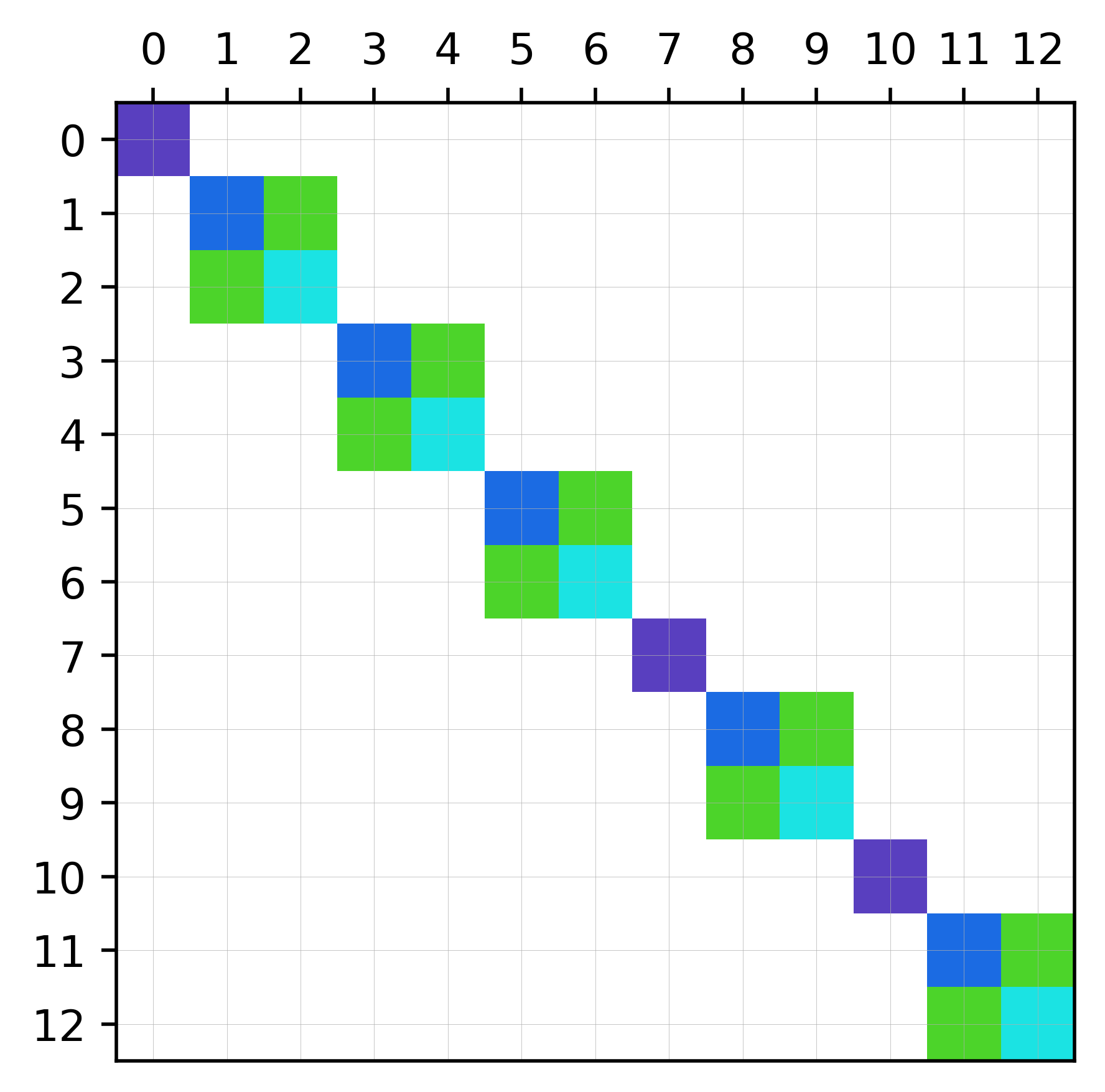}
			\caption*{$\sigma_2^{-1}$}
			\label{fig:toffoli-decomposition}
		\end{subfigure}
		\begin{subfigure}[b]{\picwidth}
			\centering
			\includegraphics[width=\textwidth]{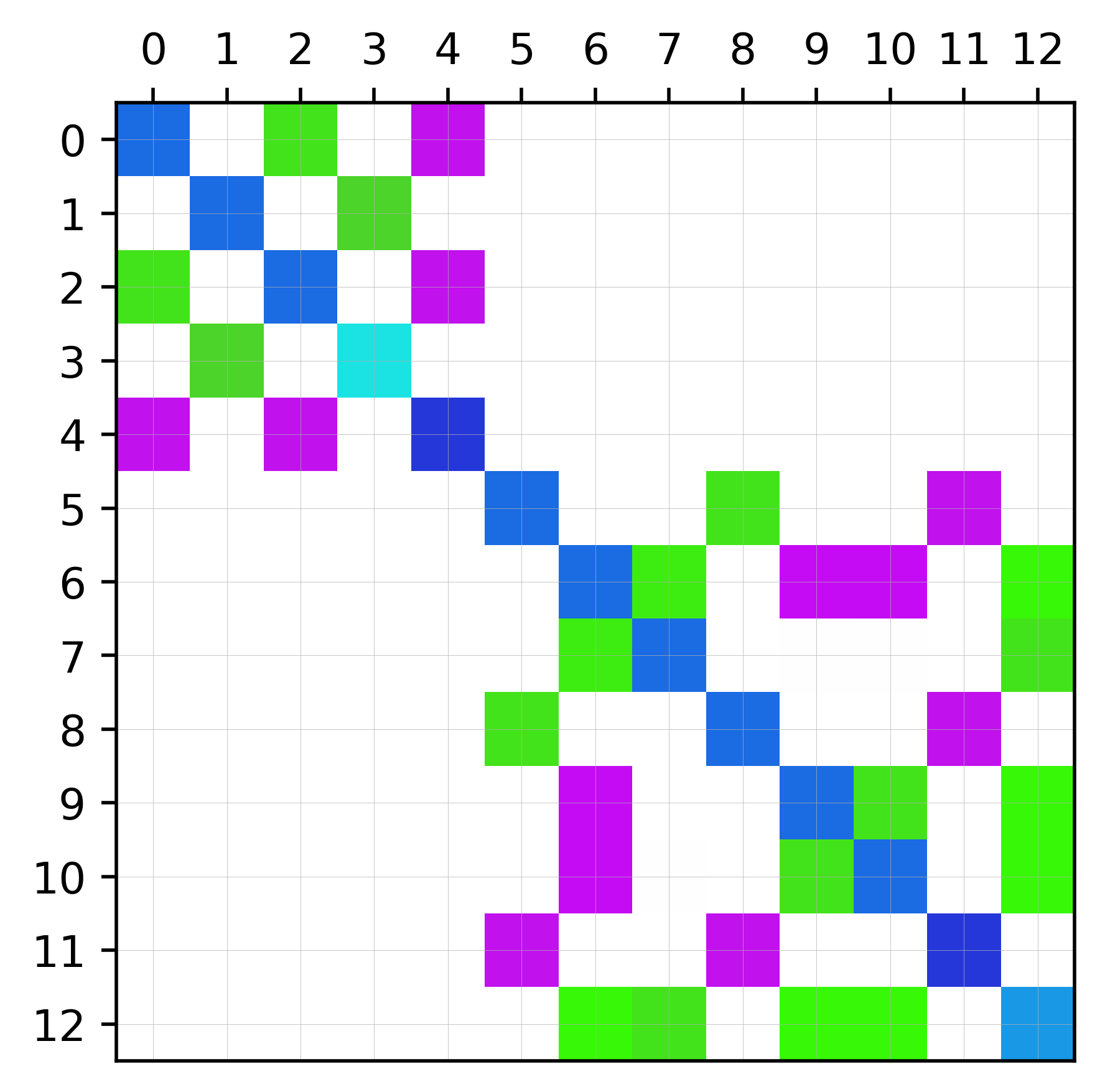}
			\caption*{$\sigma_3^{-1}$}
		\end{subfigure}
		\begin{subfigure}[b]{\picwidth}
			\centering
			\includegraphics[width=\textwidth]{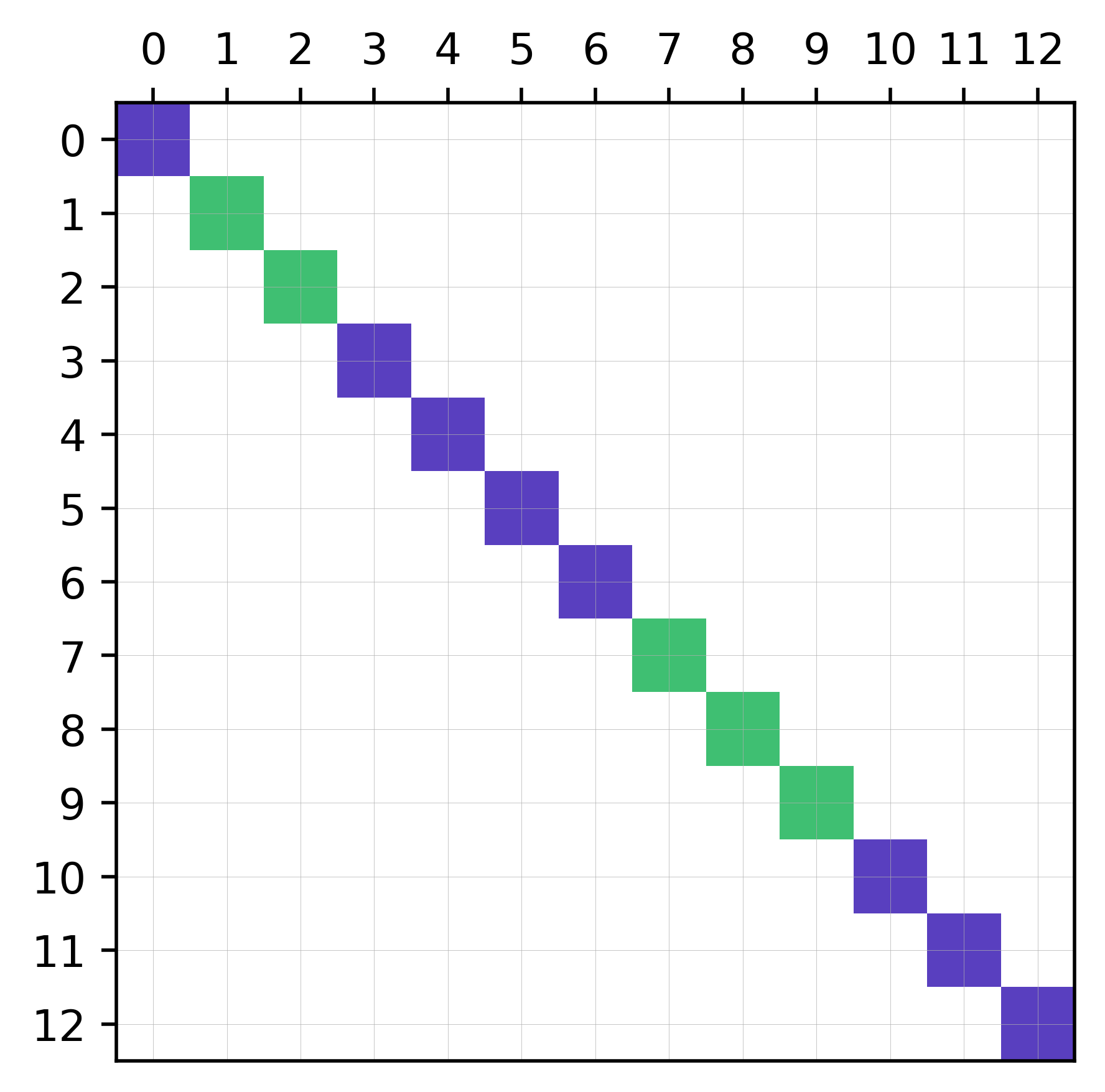}
			\caption*{$\sigma_4^{-1}$}
		\end{subfigure}
		\begin{subfigure}[b]{\picwidth}
			\centering
			\includegraphics[width=\textwidth]{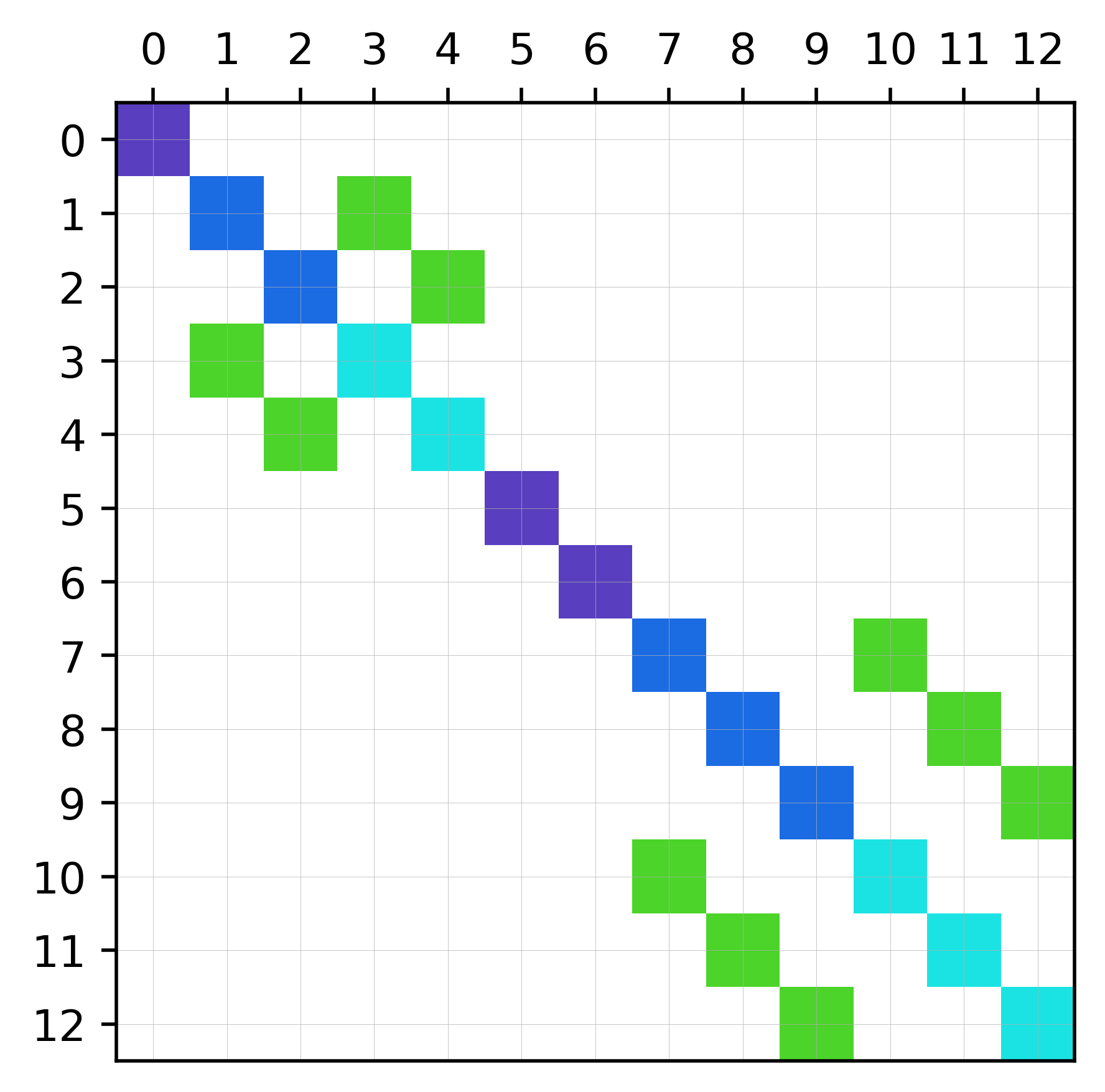}
			\caption*{$\sigma_5^{-1}$}
		\end{subfigure}
		\caption{Matrix representations of the braid generators for a system of two qubits composed of six Fibonacci anyons, {arranged in groups} of three anyons per qubit. Given six anyons, we should obtain five possible braid generators along with their inverses.
			The clockwise braid generators (up raw), {and} their respective inverses (bottom raw), are depicted using the color map scale in Fig. \ref{fig:scale}. To ensure their correctness, Artin braid group relations have been verified. The ordering of the fusion states in this figure follows the same convention as the one presented in Tab. \ref{tab:basis-2q}.}
		\label{fig:braid-generators}
	\end{figure*}
	\begin{figure*}[t]
		\centering
		\begin{subfigure}[t]{0.24\linewidth}
			\centering
			\includegraphics[width=\textwidth]{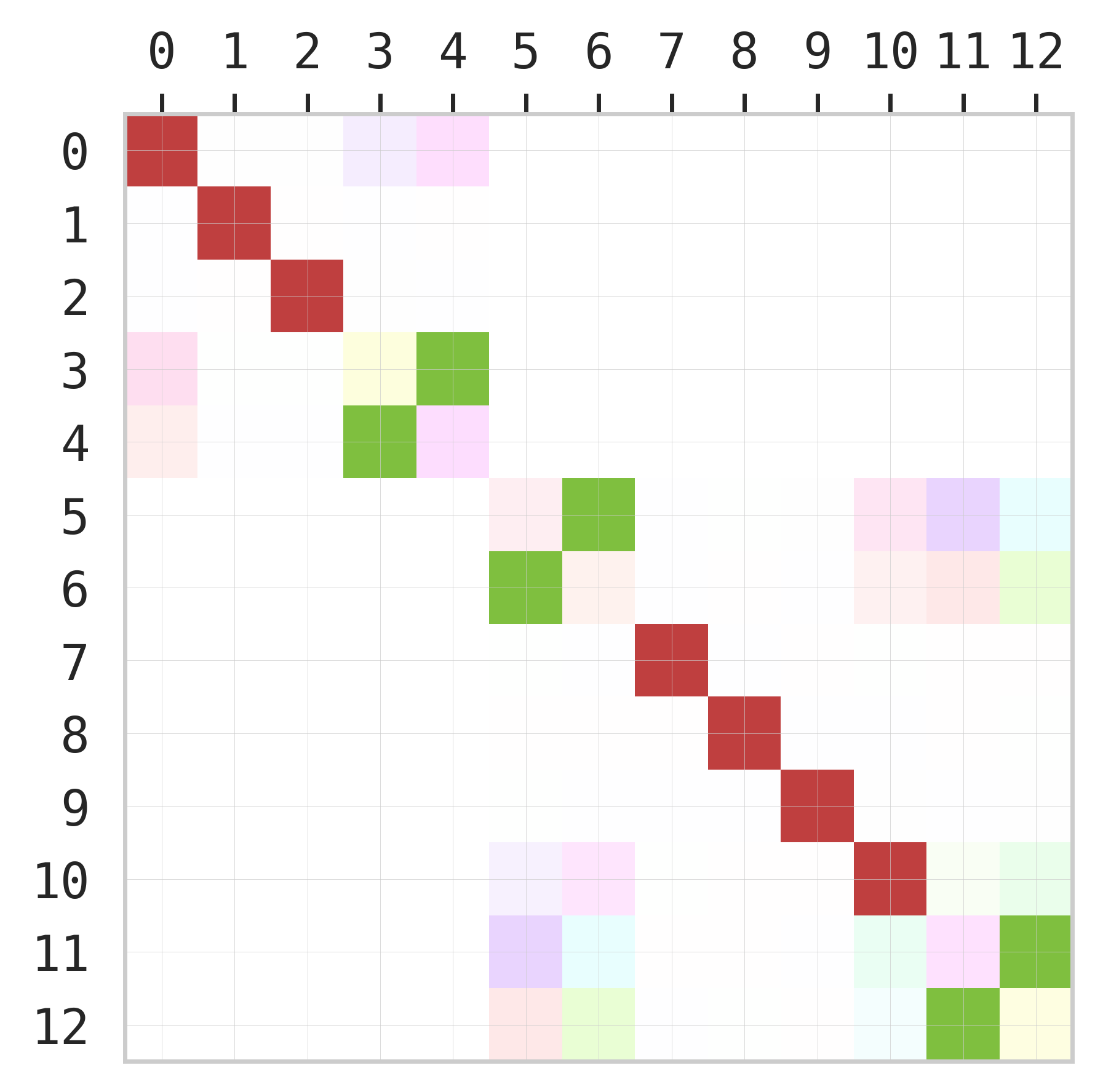}
			\caption{}
		\end{subfigure}
		\begin{subfigure}[t]{0.24\linewidth}
			\centering
			\includegraphics[width=\textwidth]{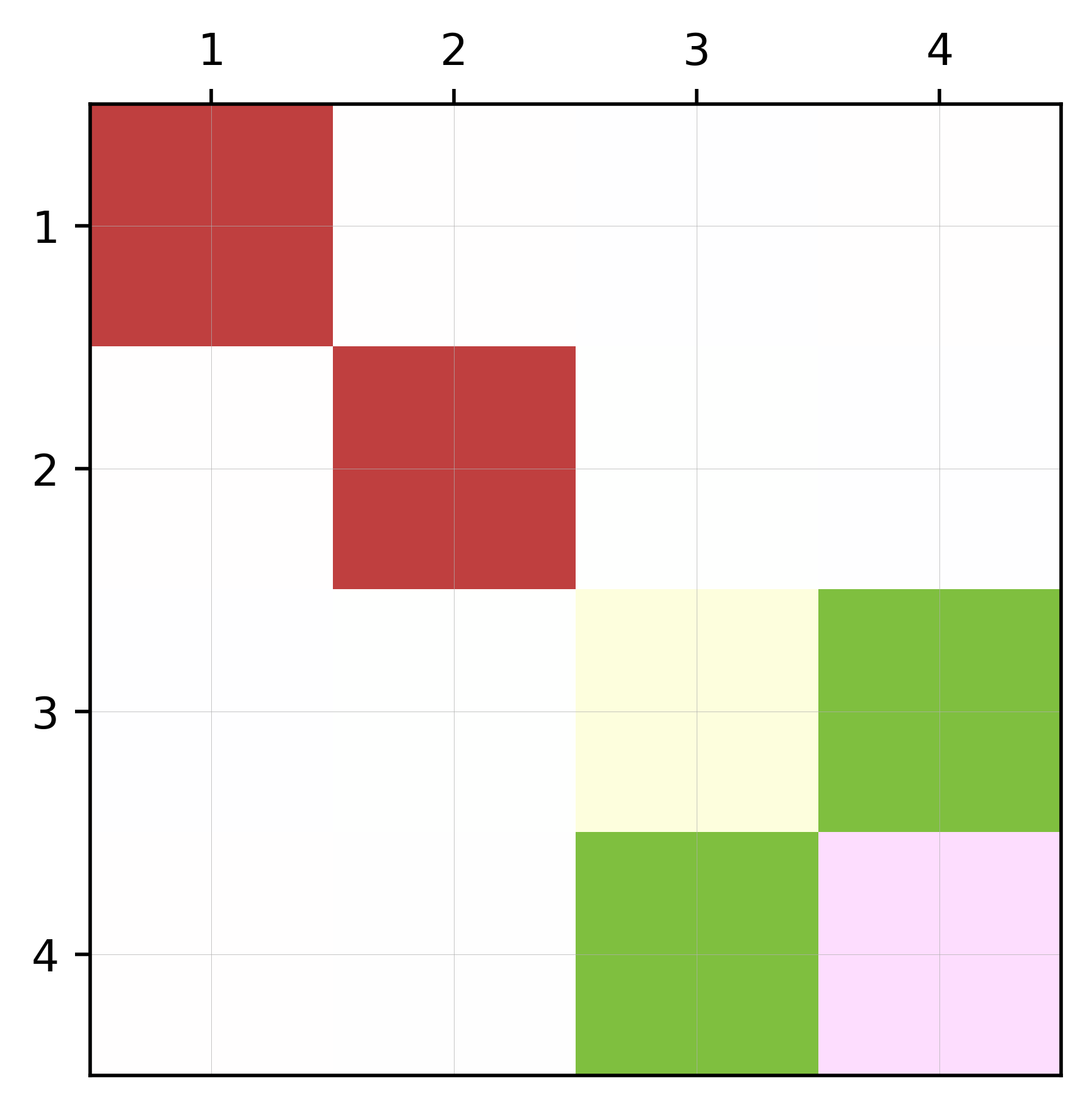}
			\caption{}
		\end{subfigure}
		\begin{subfigure}[t]{0.24\linewidth}
			\centering
			\includegraphics[width=\textwidth]{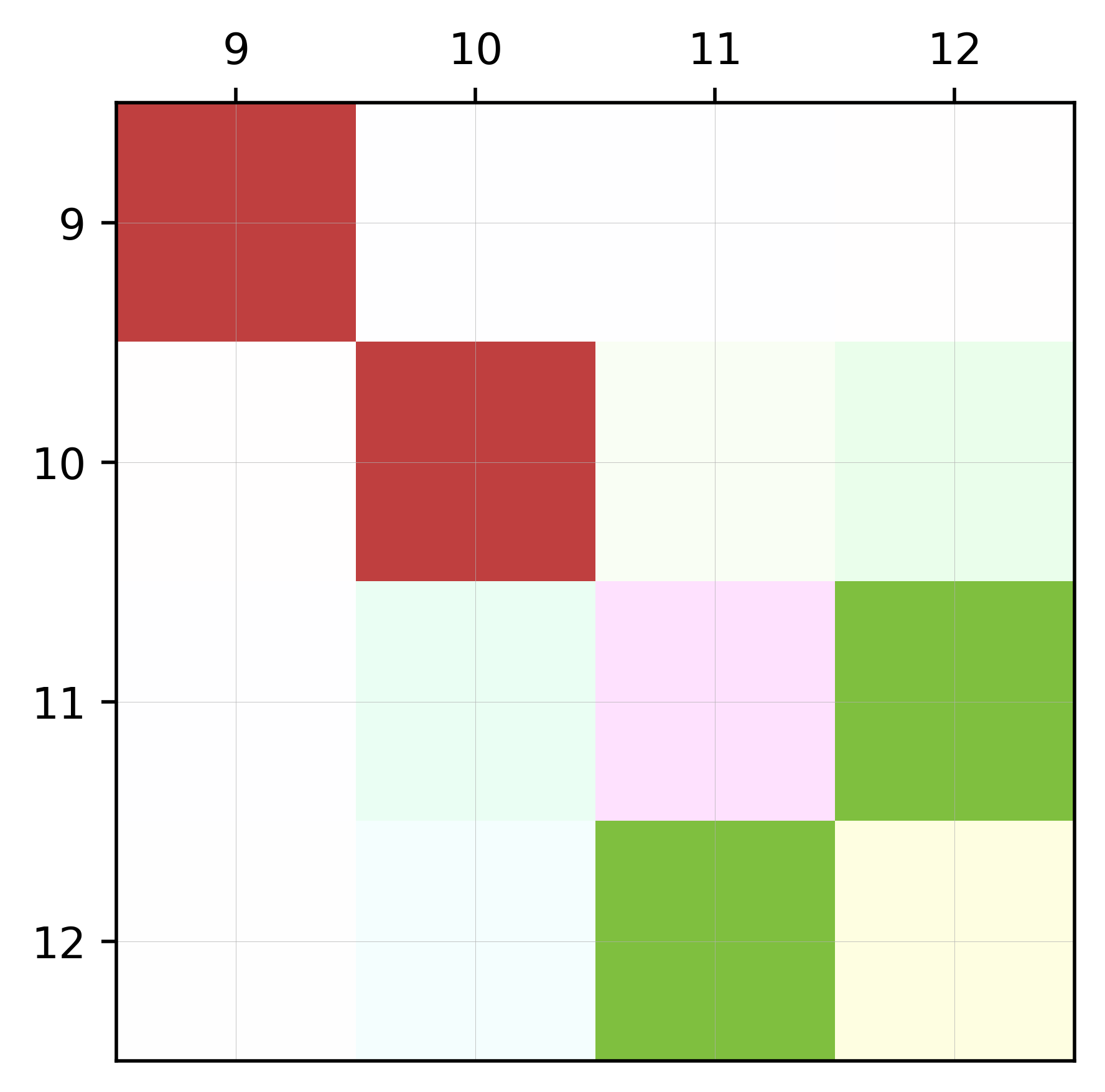}
			\caption{}
		\end{subfigure}
		\begin{subfigure}[t]{0.24\linewidth}
			\centering
			\includegraphics[width=\textwidth]{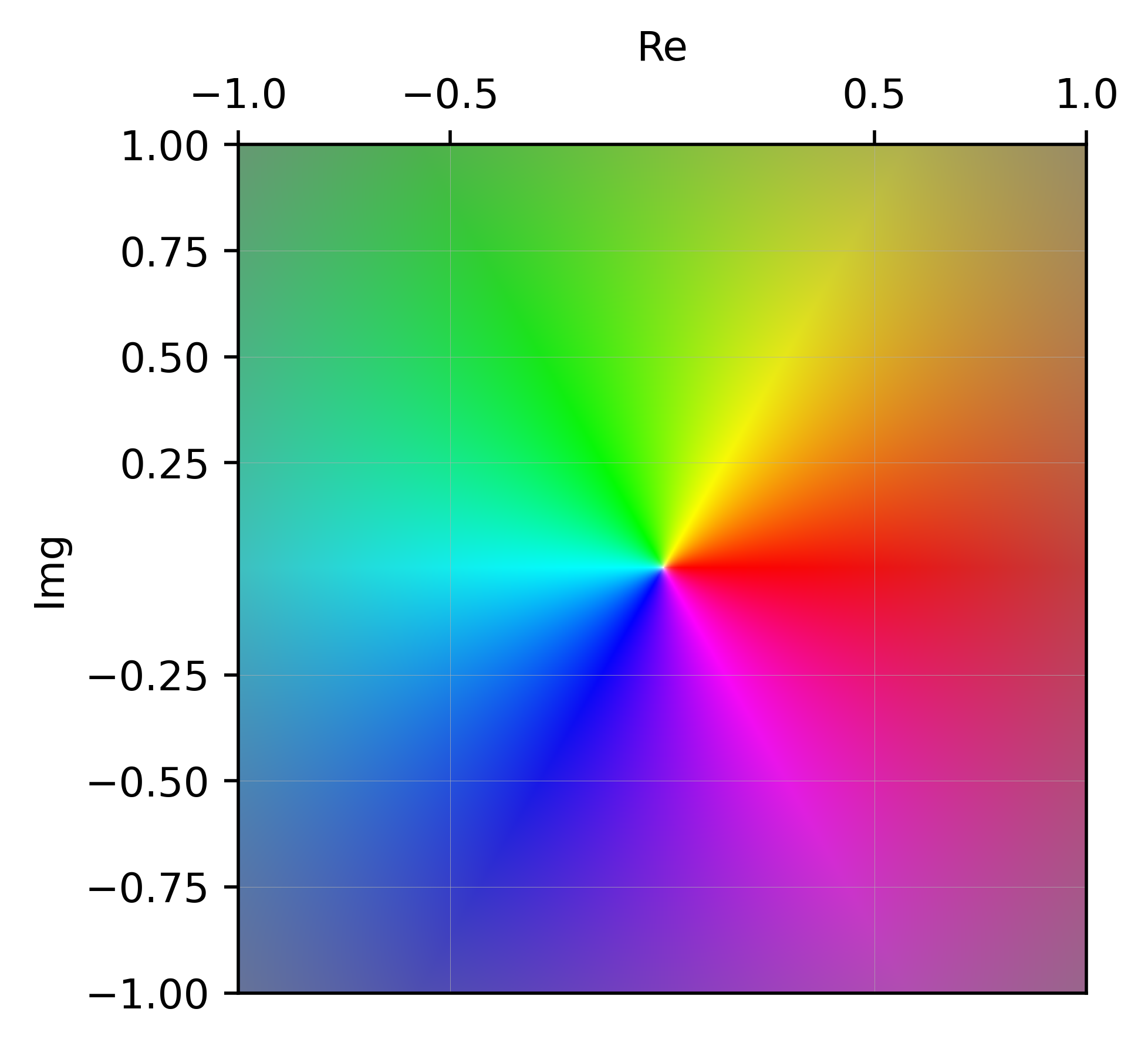}
			\caption{}
			\label{fig:scale}
		\end{subfigure}
		\caption{{(a) Matrix representation of the CNOT gate based on the braid sequence approximated by Bonesteel et al. The braid operators composing the sequence were generated using our systematic method applied on the fusion basis state of two qubits composed of three anyons each. The states are labeled in the same manner as shown in Tab. \ref{tab:basis-2q}.
				(b), (c) Submatrices corresponding to the application of the approximated CNOT braid sequence on the computational space corresponding to sector $\vac$ and sector $\fib$, respectively. The indices correspond to the computational states as illustrated in Tab. \ref{tab:basis-2q}.
				(d) Color representation of complex numbers with argument $\phi$ and modulus $r\leq 1$. The arguments are mapped linearly to the RGB spectrum, while the modulus is mapped to luminosity. The luminosity decays proportionally to $e^{-r/\sigma}$, from 1 (white) at the origin to 0.5 (saturated) at the circumference. Here, $\sigma=0.01$ is the mean radius.}
		}
		\label{fig:cnot}
	\end{figure*}
	\paragraph*{Computation of braid operators.} Since the $F$ and $R$ matrices are known for the Fibonacci model \cite{Rouabah2021}, it is possible to obtain the braid generators that operate on the above fusion basis states using the systematic method explained in the previous section. The explicit matrix representations of the braid generators are shown in Fig. \ref{fig:braid-generators}. Note that the braid operations $\sigma_1$ and $\sigma_2$ apply to the first qubit, while $\sigma_4$ and $\sigma_5$ act on the state of the second qubit. Furthermore, the braid operator $\sigma_3$ is responsible for the interaction between the two qubits. 
	However, this interaction comes at a cost, as $\sigma_3$ mixes computational states with non-computational states, inducing information leakage \cite{Ainsworth2011}. For the sake of {time} performance and {save of memory}, it is worth noting that the computation of such braid generators can be carried out {separately on the $\fib$ and $\vac$ sectors to avoid trivial components}.\\
	\paragraph*{Implementation of topological CNOT gate.} At this stage, we have attained the capability to simulate {the quantum evolution of two qubits represented by 6 Fibonacci anyons}. As an illustrative instance, we opt for a braid sequence that approximates the CNOT gate implemented with 280 successive braiding operations using an injection method \cite{Bonesteel2005}, allowing us to validate our approach.
	Computing the matrix representation of the given braid sequence results in a $13 \times 13$ matrix that includes the two possible computational sectors, as explained earlier.
	We obtain a braid-based Controlled-NOT operation, where the second qubit controls the application of the NOT gate on the first qubit, as shown in Fig. \ref{fig:cnot}. The accuracy and leakage of this approximation with respect to the \textit{conventional} CNOT quantum gate \cite{Nielsen2010} can be measured directly from the matrix representation. 
	The spectral distance $\mathcal{D}(U_1, U_2) = \sqrt{\text{maxEigenvalue}(AA^\dagger)}$ is used to measure accuracy, where $A$ is the difference between the unitaries $U_1$ and $U_2$ after eliminating the global phases \cite{Bonesteel2005}. Furthermore, leakage is measured by the quantity $1 -\sqrt{\text{minEigenvalue} (UU^\dagger)}$, where the second term is the minimum factor by which the matrix $U$ can change the norm of a quantum state \cite{Field2018}.
	The accuracy of the {topological} CNOT gate as approximated in sector $\vac$ and sector $\fib$ is {$1.73\cdot 10^{-3}$ and  $1.24\cdot 10^{-3}$, respectively, with an estimated leakage of $1.17\cdot 10^{-6}$ and $2.54\cdot 10^{-6}$}, respectively.
	The preceding example reveals that the systematic method we present in this study yields a commensurate level of accuracy as the original work \cite{Bonesteel2005}.
	
	\section{Preparation of the GHZ State}
	\label{sec:ghz}
	The primary objective of deriving a general formula for braiding anyons is to facilitate the numerical analysis of a larger array of topological qubits. In this section of the manuscript, we will examine an instance of the Greenberger-Horne-Zeilinger (GHZ) state preparation circuit. The GHZ state is renowned for its application in the GHZ gedanken experiment, which provides compelling evidence of the discord between quantum mechanics and any theory predicated on locality and reality \cite{10.1119/1.16243}.
	For our purpose, it is sufficient to use the $-i$Hadamard gate as approximated in \cite{Rouabah2021} and the controlled-$i$NOT gates as simulated in Sec. \ref{sec:cnot}. The GHZ state of $n$ qubits can be prepared by applying the $-i$Hadamard gate on the last qubit, followed by a series of consecutive C$i$NOT gates on the adjacent qubits. If the qubits are initially in the $\ket{00\cdots 0}$ state, the previous circuit results in the following state:
	\begin{align}
		\label{eq:ghz}
		\ket{\text{GHZ}} &= -\frac{i}{\sqrt{2}}\left(\ket{00\cdots 0} + i^{n-1} \ket{11\cdots 1}\right).
	\end{align}
	The first non-trivial state for which $i^{n-1} = 1$ requires five qubits. For this case, we are going to simulate the circuit and yield the states taking into consideration all possible computational sectors. Since those gates require three Fibonacci anyons per qubit with overall charge $\fib$, we distinguish eight sectors, defined by different fusion outcomes $j_1, j_2, j_3$ and $j_4$ as listed in Tab. \ref{tab:15-fusion-space} and \ref{tab:ghz-benchmark}. The numerical calculations show that the fusion space of 15 Fibonacci anyons contains 987 fusion states distributed between 16 sectors among which only 8 sectors include computaitonal states. The detailed number of the fusion space composition is elaborated in Tab. \ref{tab:15-fusion-space}.
	
	\begin{table}[t]
		\caption{The fusion space of 15 Fibonacci anyons grouped in 5 sets (qubits) accepts 987 eligible fusion states, distributed among 16 different sectors, which are the fusion states of qubits' outcomes as illustrated in Fig. \ref{fig:multi-qubit-state}. Consequently, each sector contains a set of fusion states, and eight of those sectors can accept computational states. The number of computational states equals to $2^5=32$. 
			{Note that sectors sharing the same overall charge belong to the same sub-space. Consequently, we can identify a set of braid operators that map between these sectors. Conversely, sectors with different total charges belong to separate sub-spaces}.
		}
		\label{tab:15-fusion-space}
		\begin{tabular*}{\linewidth}{@{\extracolsep\fill}lp{0.1cm}p{0.1cm}p{0.1cm}p{0.1cm}cp{2cm}@{\extracolsep\fill}}
			\toprule
			{Independent} & \multicolumn{4}{c}{Sector}    & \multirow{2}{0.5cm}{States}  & \multirow{2}{0.5cm}{\centering Computational}\\
			\cmidrule{2-5}
			Sub-spaces    & $j_1$ & $j_2$ & $j_3$ & $j_4$ & & \\
			\midrule
			\multirow{8}*{Sub-space $\vac$}
			& $\vac$ & $\vac$ & $\vac$ & $\vac$ & $5$   & No  \\
			& $\fib$ & $\vac$ & $\vac$ & $\vac$ & $16$  & No  \\
			& $\vac$ & $\fib$ & $\vac$ & $\vac$ & $20$  & No  \\
			& $\fib$ & $\fib$ & $\vac$ & $\vac$ & $48$  & No  \\
			& $\vac$ & $\vac$ & $\fib$ & $\vac$ & $20$  & No  \\
			& $\fib$ & $\vac$ & $\fib$ & $\vac$ & $64$  & Yes \\
			& $\vac$ & $\fib$ & $\fib$ & $\vac$ & $60$  & Yes \\
			& $\fib$ & $\fib$ & $\fib$ & $\vac$ & $144$ & Yes \\
			\midrule
			\multirow{8}*{Sub-space $\fib$}
			& $\vac$ & $\vac$ & $\vac$ & $\fib$ & $10$  & No  \\
			& $\fib$ & $\vac$ & $\vac$ & $\fib$ & $32$  & No  \\
			& $\vac$ & $\fib$ & $\vac$ & $\fib$ & $40$  & Yes \\
			& $\fib$ & $\fib$ & $\vac$ & $\fib$ & $96$  & Yes \\
			& $\vac$ & $\vac$ & $\fib$ & $\fib$ & $30$  & No  \\
			& $\fib$ & $\vac$ & $\fib$ & $\fib$ & $96$  & Yes \\
			& $\vac$ & $\fib$ & $\fib$ & $\fib$ & $90$  & Yes \\
			& $\fib$ & $\fib$ & $\fib$ & $\fib$ & $216$ & Yes \\
			\bottomrule
		\end{tabular*}
	\end{table}
	
	To assess the quality of the GHZ state generated by this approximation, we employ two methodologies. Firstly, we analyze the degree of the logical entanglement established among qubits, comparing it with the ideal scenario. It is important to distiguish between the logical entanglement that takes into consideration the computational basis only and not the anyonic entanglement which someone can compute with the anyonic von Neumann entropy \cite{bonderson2017}. Secondly, we compute the spectral distance between the GHZ state and its approximated counterpart.
	To quantify the logical entanglement, we compute the von Neumann entropy of entanglement for each qubit $q$, defined as:
	\begin{align}
		S_q(\rho) = -\tr(\rho_q \ln \rho_q)
	\end{align}
	such that $\rho$ is the density matrix of the qubits system, and $\rho_q = \tr_{(q^c)} \rho$ is the partial trace of $\rho$ after tracing over $q^c$ the set of qubits that complement the qubit $q$. The exact entanglement entropy of each qubit in the GHZ state should equal to $\ln{2}$. In Tab \ref{tab:ghz-benchmark}, we list the difference $\ln{2} - S_q(\rho)$, for each $q=1,\cdots,5$ and each computational sector. {In summary, the distinction lies in the order of magnitude, typically around} $10^{-4}$.
	
	On the other hand, the squared inner product between the GHZ state \eqref{eq:ghz} and the approximated one {is computed to give the approximation inaccuracy}. This operation yields the error values listed in Tab. \ref{tab:ghz-benchmark}. This shows that the prepared GHZ state should be in theory close to the ideal GHZ state up to some error of $5.785\cdot 10^{-5}$ in average. It is important also to quantify the probability of measuring non-computational state at the end. This is the leakage amount that can be calculated by extracting the state's norm from 1. We can check in Tab. \ref{tab:ghz-benchmark} the leakage values in different sectors varying around the mean $7.153\cdot 10^{-6}$.
	Before concluding this section, it is interesting to notice the trade-off between the errorness of the state and the leakage.
	{This observation is pertinent because its cause lies in the injection braid, which is primarily responsible for leakage. The injection braid typically involves weaving where only one anyon is moved around other anyons, approximating the identity gate \cite{Bonesteel2005, Simon2006}. Injection is employed to efficiently construct controlled gates with anyons \cite{Bonesteel2005, Hormozi2007, tounsi2023optimized}. In CNOT gates, we utilize injection weaves consisting of three strands. Generally, an injection braid $I$ transforms the initial state of three anyons into a similar state as follows:
		\begin{align}
			I \ket{((a, b)_i, c)_j} = c_{ii} \ket{((a, b)_i, c)_j} + \sum_{k\neq i} c_{ik} \ket{((a, b)_k, c)_j}
		\end{align}
		where $|c_{ii}|^2 \approx 1$. It is natural for $c_{ik}$ to vary for different $i$ values. Consequently, when the injection braid is employed to inject an anyon between two qubits, different $i$ values are obtained with varying probability amplitudes in the available topological sectors, resulting in different leakage values. However, a large leakage value in a specific sector absorbs out the unwanted logical states of amplitudes $c_{ik}$ such that $k\neq i$, thus leading to better accuracy, and vice versa. This analysis holds true for any method utilizing the injection braid.
	}
	
	\begin{table*}[t]
		\caption{Comparative analysis between ideal and simulated GHZ states of five qubits. The employment of three Fibonacci anyons per qubit infers the existence of eight computational sectors, represented as $[j_1, j_2, j_3, j_4]$. This is contingent upon the premise that the computational states of each three anyon qubit maintain an aggregate charge of $\fib$. The logical entanglement entropy difference indicates the entropy extent to which the approximated state deviate from the maximal entanglement of the GHZ state. The column of errors describes the distance between the GHZ state and the approximated state in each sector. Lastly, the leakage column quantifies the amount of non-computational states leaked to the approximated state.}
		\label{tab:ghz-benchmark}
		\begin{tabular*}{\linewidth}{@{\extracolsep\fill}cccc|p{1cm}p{1cm}p{1cm}p{1cm}p{1cm}|c|c}
			\toprule
			\multicolumn{4}{c|}{Sector} & \multicolumn{5}{c|}{Logical Entanglement Entropy Difference ($\times 10^{-5}$)} & {Error} & {Leakage}\\
			\midrule
			$j_1$ & $j_2$ & $j_3$ & $j_4$ & \centering Qubit 1 & \centering Qubit 2 & \centering Qubit 3 & \centering Qubit 4 & \centering Qubit 5 & ($\times 10^{-5}$) & ($\times 10^{-6}$) \tabularnewline
			\midrule
			$\fib$ & $\vac$ & $\fib$ & $\vac$ & \centering$3.168$ &\centering$3.068$ &\centering$3.068$ &\centering$3.069$ &\centering$3.144$ & $5.699$ & $6.799$ \\
			$\vac$ & $\fib$ & $\fib$ & $\vac$ & \centering$3.140$ &\centering$3.067$ &\centering$3.068$ &\centering$3.069$ &\centering$3.170$ & $5.699$ & $6.799$ \\
			$\fib$ & $\fib$ & $\fib$ & $\vac$ & \centering$3.109$ &\centering$3.036$ &\centering$3.036$ &\centering$3.037$ &\centering$3.111$ & $5.642$ & $7.846$ \\
			$\vac$ & $\fib$ & $\vac$ & $\fib$ & \centering$3.144$ &\centering$3.123$ &\centering$3.124$ &\centering$3.124$ &\centering$3.226$ & $6.461$ & $4.921$ \\
			$\fib$ & $\fib$ & $\vac$ & $\fib$ & \centering$3.043$ &\centering$3.022$ &\centering$3.023$ &\centering$3.023$ &\centering$3.098$ & $5.530$ & $8.294$ \\
			$\fib$ & $\vac$ & $\fib$ & $\fib$ & \centering$3.083$ &\centering$3.010$ &\centering$3.010$ &\centering$3.011$ &\centering$3.086$ & $5.242$ & $8.721$ \\
			$\vac$ & $\fib$ & $\fib$ & $\fib$ & \centering$3.106$ &\centering$3.079$ &\centering$3.080$ &\centering$3.081$ &\centering$3.182$ & $6.031$ & $6.381$ \\
			$\fib$ & $\fib$ & $\fib$ & $\fib$ & \centering$3.074$ &\centering$3.047$ &\centering$3.047$ &\centering$3.048$ &\centering$3.122$ & $5.971$ & $7.464$ \\
			\bottomrule
		\end{tabular*}
	\end{table*}
	
	\section{Conclusion}
	
	Quantum information can be encoded, manipulated, and protected in a topologically robust manner, by performing braiding operations.
	{In this paper, we presented a practical method simplifying the systematic computation of braid operators indispensable for the compilation of quantum circuits. The significance of our study lies in its relevance to {any arbitrary fusion order of anyons including the sparse encoding for} qubit or qudit-based topological quantum computation} systems that utilize anyonic states  \cite{Rouabah2021}. 
	Importantly, our method {exhibits systematic extendibility} to encompass all $SU(2)_k$ anyon models and others by {computing} the $F$ and $R$ matrices through their algebraic expressions \cite{Kohno1990, bonderson, Genetay2021} or the solution of consistency equations.
	Moreover, we validate the introduced formalism through numerical verification using algebraic braid relations. Furthermore, {we study} a previously established topological CNOT gate compiled with braiding Fibonacci anyons.
	Subsequently, {the CNOT gate along with the Hadamard gate} have been utilized to prepare the GHZ state of five qubits, or equivalently, fifteen anyons in total. The computational sectors in this scenario have been numerically determined. The quality of the GHZ state approximation has been evaluated using two metrics: the logical entanglement entropy and the inner product distance. Both metrics yield approximations in the order of $10^{-4}$. This suggests a reasonable degree of accuracy in the GHZ state approximation in comparison to the building gates accuracies, indicating the effectiveness of the simulation and the potential for further exploration in this field.
	The method {detailed} in this paper has been {proficiently} implemented for the {simulation of topological quantum circuits featuring} Fibonacci {anyons within the} open-source numerical library {named TQSim  \cite{TQSim}, developed by the authors of this paper. {Furthermore, this approach has been successfully applied to implement novel optimized three-qubit controlled gates in the Fibonacci anyon model} \cite{tounsi2023optimized}.
		In conclusion, {the application of this method in simulating topological quantum computation is highly valuable, serving not only for testing quantum gate compilation but also for the development of topologically protected quantum circuits}.

	\section*{Acknowledgements}
		This document has been produced with the financial assistance of the European Union (Grant no. DCI-PANAF/2020/420-028), through the African Research Initiative for Scientific Excellence (ARISE), pilot programme. ARISE is implemented by the African Academy of Sciences with support from the European Commission and the African Union Commission.
		We are grateful to the Algerian Ministry of Higher Education and Scientific Research and DGRST for the financial support.

	\bibliography{bibliography}
	\bibliographystyle{abbrv}

\appendix
		
		\section{The General Knitting Matrix Components}
		\label{app:supplementary-materials}
		The general form of the knitting matrices $K$ and $L$, defined for a fusion state of an arbitrary number of qudits with identical $N=q+1$ anyons per qudit. The $K$ matrix describes the effect of braiding two adjacent anyons shared between the $m$'th and the {$(m+1)$th} neighboring qudits. On the other hand, braiding two adjacent anyons within the same qudit is already described by the braid matrix $B$ as explained in Sec. \ref{sec:braiding-matrix-B}. First, to define the knitting matrix, the $L$ braiding move is defined in Eq. \eqref{eq:general-L} and its general formula is elaborated in Eq. \eqref{eq:general-L-definition}. Finally, the knitting move, the general action of the braid operator $\sigma_n$ on an arbitrary fusion state such that $n=mN$, is defined generally in Eq. \eqref{eq:general-K} while its formula is shown in Eq. \eqref{eq:general-K-definition}.
		
		\def\unitx{2.66}
		\def\unity{3.33}
		\def\minsize{25}
		\def\nq{2} 
		\def\na{4} 
		\def\den{4} 
		\def\versize{0.33\linewidth}

		\def\Lstatei{
			\resizebox{\versize}{!}{
				\tikz[baseline=.1ex,
				roundnode/.style={circle, draw=blue!60, fill=blue!5, minimum size=\minsize mm},
				anyonnode/.style={circle, draw=red!60, fill=red!5, thick, minimum size=\minsize mm},
				]{
					\def\orgnx{0};
					\def\orgny{0};
					
					\pgfmathsetmacro\nr{\na-1};
					\pgfmathsetmacro\nrq{\nq-1};
					
					\foreach \q in {0,...,\nrq}{
						\foreach \a in {1,...,\nr}{
							\def\rootx{\orgnx + \q * \unitx * \na + \a * \unitx/2};
							\def\rooty{\orgny -\unity/2 * \a};
							\pgfmathsetmacro\temp{\nq-2};
							\ifthenelse{\a=2}{
								\draw[ultra thick, loosely dotted, black]  (\rootx, \rooty) -- 
								(\rootx - \unitx/2, \rooty + \unity/2) ;
							}{
								\draw[ultra thick]  (\rootx, \rooty) -- 
								(\rootx - \unitx/2, \rooty + \unity/2) ;
							}
							\draw[ultra thick] (\rootx, \rooty)
							-- (\rootx + \a * \unitx/2, \orgny);
						};
					};
					
					\pgfmathsetmacro\orgnxr{\orgnx + \nr * \unitx/2};
					\pgfmathsetmacro\orgnyr{\orgny -\unity/2 * \nr};
					\pgfmathsetmacro\nrr{\nq-1};
					\pgfmathsetmacro\unitxr{\unitx * \na};
					\pgfmathsetmacro\unityr{\unity * \na};
					
					\foreach \a in {1,...,\nrr}{
						\def\rootx{\orgnxr + \a * \unitxr/2};
						\def\rooty{\orgnyr -\unityr/2 * \a};
						\draw[ultra thick]  (\rootx, \rooty)
						-- (\rootx - \unitxr/2, \rooty + \unityr/2);
						\draw[ultra thick] (\rootx, \rooty)
						-- (\rootx + \a * \unitxr/2, \orgnyr);
					};
					
					\def\xa{\orgnx + 3 * \unitx};
					\def\ya{\orgny};
					\def\xb{\orgnx + \unitx + 3 * \unitx};
					\def\yb{\orgny};
					\draw (\xb,\yb) .. controls (\xb, \yb+\unity/2) and (\xa,\ya+\unity/2) .. (\xa,\ya+\unity);
					\fill[white] (\xa+\unitx/2,\ya+\unity/2) circle (\unitx/10);
					\draw (\xa,\ya) .. controls (\xa,\ya+\unity/2) and (\xb,\yb+\unity/2) .. (\xb,\yb+\unity);
					
					\foreach \q in {0,...,\nrq}{
						\foreach \a in {0,...,\nr}{
							\def\x{\orgnx + \q * \unitx * \na + \a * \unitx};
							\def\y{\orgny};
							\ifthenelse{\q = 0 \AND \a = 3}{
								\filldraw[black] (\x, \y+\unity) circle (0pt) node[anyonnode]{\Huge $a^{m+1}_{0}$}; 
							}{
								\ifthenelse{\q = 1 \AND \a = 0}{
									\filldraw[black] (\x, \y+\unity) circle (0pt) node[anyonnode]{\Huge $a^m_q$}; 
								}{
									\ifthenelse{\q=1}{
										\ifthenelse{\a=3}{
											\filldraw[black] (\x, \y) circle (0pt) node[anyonnode]{\Huge $a^{m+1}_{q}$};
										}{
											\ifthenelse{\a=2}{
												\filldraw[black] (\x, \y) circle (0pt) node[anyonnode]{\Huge $a^{m+1}_{q-1}$};
											}{
												\filldraw[black] (\x, \y) circle (0pt) node[anyonnode]{\Huge $a^{m+1}_{\a}$};
											};
										};
									}{
										\ifthenelse{\a=2}{
											\filldraw[black] (\x, \y) circle (0pt) node[anyonnode]{\Huge $a^m_{q-1}$};
										}{
											\filldraw[black] (\x, \y) circle (0pt) node[anyonnode]{\Huge $a^m_{\a}$};
										};
									};
								};
							};
						};
					};
					
					\foreach \q in {1,...,\nq}{
						\foreach \a in {1,...,\nr}{
							\pgfmathsetmacro\p{\q - 1};
							\def\rootx{\orgnx + \p * \unitx * \na + \a * \unitx/2};
							\def\rooty{\orgny -\unity/2 * \a};
							\ifthenelse{\a=\nr}{
								\ifthenelse{\q=\nq}{
									\filldraw[black] (\rootx-\unitx/\den, \rooty) circle (0pt) node[anchor=east]{\Huge $i_{(m+1)q}$};
								}{
									\filldraw[black] (\rootx-\unitx/\den, \rooty) circle (0pt) node[anchor=east]{\Huge $i_{mq}$};
								}
							}{
								\ifthenelse{\a=2}{	
									\ifthenelse{\q=\nq}{
										\filldraw[black] (\rootx-\unitx/\den, \rooty) circle (0pt) node[anchor=east]{\Huge $i_{(m+1)(q-1)}$};
									}{
										\filldraw[black] (\rootx-\unitx/\den, \rooty) circle (0pt) node[anchor=east]{\Huge $i_{m(q-1)}$};
									}
								}{
									\ifthenelse{\a=1}{	
										\ifthenelse{\q=\nq}{
											\filldraw[black] (\rootx-\unitx/\den, \rooty) circle (0pt) node[anchor=east]{\Huge $i_{(m+1)1}$};
										}{
											\filldraw[black] (\rootx-\unitx/\den, \rooty) circle (0pt) node[anchor=east]{\Huge $i_{m1}$};
										}
							}}}
						};
					};
					
					\foreach \a in {1,...,\nrr}{
						\def\rootx{\orgnxr + \a * \unitxr/2};
						\def\rooty{\orgnyr -\unityr/2 * \a};
						\ifthenelse{\a=\nrr}{
							\filldraw[black] (\rootx, \rooty) circle (0pt) node[roundnode]{\Huge $k$};
						}{
							\filldraw[black] (\rootx, \rooty) circle (0pt) node[roundnode]{\Huge $j_{\a}$};
						}
					};
				} 
			} 
		} 
		
		\def\Lstatef{
			\resizebox{\versize}{!}{
				\tikz[baseline=.1ex,
				roundnode/.style={circle, draw=blue!60, fill=blue!5, minimum size=\minsize mm},
				anyonnode/.style={circle, draw=red!60, fill=red!5, thick, minimum size=\minsize mm},
				]{
					\def\orgnx{0};
					\def\orgny{0};
					
					\pgfmathsetmacro\nr{\na-1};
					\pgfmathsetmacro\nrq{\nq-1};
					
					\foreach \q in {0,...,\nrq}{
						\foreach \a in {1,...,\nr}{
							\def\rootx{\orgnx + \q * \unitx * \na + \a * \unitx/2};
							\def\rooty{\orgny -\unity/2 * \a};
							\pgfmathsetmacro\temp{\nq-2};
							\ifthenelse{\a=2}{
								\draw[ultra thick, loosely dotted, black]  (\rootx, \rooty) -- 
								(\rootx - \unitx/2, \rooty + \unity/2) ;
							}{
								\draw[ultra thick]  (\rootx, \rooty) -- 
								(\rootx - \unitx/2, \rooty + \unity/2) ;
							}
							\draw[ultra thick] (\rootx, \rooty)
							-- (\rootx + \a * \unitx/2, \orgny);
						};
					};
					
					\pgfmathsetmacro\orgnxr{\orgnx + \nr * \unitx/2};
					\pgfmathsetmacro\orgnyr{\orgny -\unity/2 * \nr};
					\pgfmathsetmacro\nrr{\nq-1};
					\pgfmathsetmacro\unitxr{\unitx * \na};
					\pgfmathsetmacro\unityr{\unity * \na};
					
					\foreach \a in {1,...,\nrr}{
						\def\rootx{\orgnxr + \a * \unitxr/2};
						\def\rooty{\orgnyr -\unityr/2 * \a};
						\draw[ultra thick]  (\rootx, \rooty)
						-- (\rootx - \unitxr/2, \rooty + \unityr/2);
						\draw[ultra thick] (\rootx, \rooty)
						-- (\rootx + \a * \unitxr/2, \orgnyr);
					};
					\foreach \q in {0,...,\nrq}{
						\foreach \a in {0,...,\nr}{
							\def\x{\orgnx + \q * \unitx * \na + \a * \unitx};
							\def\y{\orgny};
							\ifthenelse{\q = 0 \AND \a = 3}{
								\filldraw[black] (\x, \y) circle (0pt) node[anyonnode]{\Huge $a^{m+1}_{0}$}; 
							}{
								\ifthenelse{\q = 1 \AND \a = 0}{
									\filldraw[black] (\x, \y) circle (0pt) node[anyonnode]{\Huge $a^m_q$}; 
								}{
									\ifthenelse{\q=1}{
										\ifthenelse{\a=3}{
											\filldraw[black] (\x, \y) circle (0pt) node[anyonnode]{\Huge $a^{m+1}_{q}$};
										}{
											\ifthenelse{\a=2}{
												\filldraw[black] (\x, \y) circle (0pt) node[anyonnode]{\Huge $a^{m+1}_{q-1}$};
											}{
												\filldraw[black] (\x, \y) circle (0pt) node[anyonnode]{\Huge $a^{m+1}_{\a}$};
											};
										};
									}{
										\ifthenelse{\a=2}{
											\filldraw[black] (\x, \y) circle (0pt) node[anyonnode]{\Huge $a^m_{q-1}$};
										}{
											\filldraw[black] (\x, \y) circle (0pt) node[anyonnode]{\Huge $a^m_{\a}$};
										};
									};
								};
							};
						};
					};
					
					\foreach \q in {1,...,\nq}{
						\foreach \a in {1,...,\nr}{
							\pgfmathsetmacro\p{\q - 1};
							\def\rootx{\orgnx + \p * \unitx * \na + \a * \unitx/2};
							\def\rooty{\orgny -\unity/2 * \a};
							\ifthenelse{\a=\nr}{
								\ifthenelse{\q=\nq}{
									\filldraw[black] (\rootx-\unitx/\den, \rooty) circle (0pt) node[anchor=east]{\Huge $i'_{(m+1)q}$};
								}{
									\filldraw[black] (\rootx-\unitx/\den, \rooty) circle (0pt) node[anchor=east]{\Huge $i'_{mq}$};
								}
							}{
								\ifthenelse{\a=2}{	
									\ifthenelse{\q=\nq}{
										\filldraw[black] (\rootx-\unitx/\den, \rooty) circle (0pt) node[anchor=east]{\Huge $i'_{(m+1)(q-1)}$};
									}{
										\filldraw[black] (\rootx-\unitx/\den, \rooty) circle (0pt) node[anchor=east]{\Huge $i_{m(q-1)}$};
									}
								}{
									\ifthenelse{\a=1}{	
										\ifthenelse{\q=\nq}{
											\filldraw[black] (\rootx-\unitx/\den, \rooty) circle (0pt) node[anchor=east]{\Huge $i'_{(m+1)1}$};
										}{
											\filldraw[black] (\rootx-\unitx/\den, \rooty) circle (0pt) node[anchor=east]{\Huge $i_{m1}$};
										}
							}}}
						};
					};
					
					\foreach \a in {1,...,\nrr}{
						\def\rootx{\orgnxr + \a * \unitxr/2};
						\def\rooty{\orgnyr -\unityr/2 * \a};
						\ifthenelse{\a=\nrr}{
							\filldraw[black] (\rootx, \rooty) circle (0pt) node[roundnode]{\Huge $k$};
						}{
							\filldraw[black] (\rootx, \rooty) circle (0pt) node[roundnode]{\Huge $j_{\a}$};
						}
					};
				} 
			} 
		} 
			\begin{align}
				\label{eq:general-L}
				&
				\begin{array}{c}
					\Lstatei
				\end{array}
				=\nonumber\\&
				\sum_{i'_{mq} i'_{(m+1)1} \cdots i'_{(m+1)q}}
				\left(L^{k}_{i_{m(q-1)}}\right)^{i_{mq} \cdots i_{(m+1)q}}_{i'_{mq} \cdots i'_{(m+1)q}}
				\begin{array}{c}
					\Lstatef
				\end{array}
			\end{align}
			such that:
			\begin{align}
				\label{eq:general-L-definition}
				&\left(L^{p_{q+1}}_{i_{m(q-1)}}\right)^{i_{mq} \cdots i_{(m+1)q}}_{i'_{mq} \cdots i'_{(m+1)q}} =
				\nonumber\\
				&\sum_{p_1 \cdots p_q i'_{mq}i'_{(m+1)1} i'_{(m+1)2} \cdots i'_{(m+1)q}}
				\prod_{r=1}^{q} \left(F_{i_{mq}i_{(m+1)(q-r)}a_{(m+1)(q-r+1)}}^{\dagger p_{q-r+2}} \right)^{i_{(m+1)(q-r+1)}}_{p_{q-r+1}}
				\nonumber\\
				&\left( B_{i_{m(q-1)}a_{mq}a_{(m+1)0}}^{p_1} \right)^{i_{mq}}_{i'_{mq}}
				\prod_{r=1}^{q} \left(F_{i'_{mq}i'_{(m+1)(q-r)}a_{(m+1)(q-r+1)}}^{p_{q-r+2}} \right)_{i'_{(m+1)(q-r+1)}}^{p_{q-r+1}}.
			\end{align}
			where \quad $p_{q+1} = k$, \quad $i'_{(m+1)0} = a_{mq}$ and $a_{ij}=a^i_j$.\\
			\\
			Consequently, the general action of the braid operator $\sigma_n$ on an arbitrary fusion state such as $n=m\times N$ is given as follows:
			\def\unitx{2.66}
			\def\unity{3.33}
			\def\minsize{40}
			\def\ratio{0.62} 
			\def\nq{2} 
			\def\na{4} 
			\def\den{4} 
			\def\versize{0.35\linewidth}
			
			\def\Kstatei{
				\resizebox{\versize}{!}{
					\tikz[baseline=.1ex,
					roundnode/.style={circle, draw=blue!60, fill=blue!5, minimum size=\minsize mm},
					anyonnode/.style={circle, draw=red!60, fill=red!5, thick, minimum size=\ratio*\minsize mm},
					]{
						\def\orgnx{0};
						\def\orgny{0};
						
						\pgfmathsetmacro\nr{\na-1};
						\pgfmathsetmacro\nrq{\nq-1};
						
						\foreach \q in {0,...,\nrq}{
							\foreach \a in {1,...,\nr}{
								\def\rootx{\orgnx + \q * \unitx * \na + \a * \unitx/2};
								\def\rooty{\orgny -\unity/2 * \a};
								\pgfmathsetmacro\temp{\nq-2};
								\ifthenelse{\a=2}{
									\draw[ultra thick, loosely dotted, black]  (\rootx, \rooty) -- 
									(\rootx - \unitx/2, \rooty + \unity/2) ;
								}{
									\draw[ultra thick]  (\rootx, \rooty) -- 
									(\rootx - \unitx/2, \rooty + \unity/2) ;
								}
								\draw[ultra thick] (\rootx, \rooty)
								-- (\rootx + \a * \unitx/2, \orgny);
							};
						};
						
						\pgfmathsetmacro\orgnxr{\orgnx -  \unitx * \na + \nr * \unitx/2};
						\pgfmathsetmacro\orgnyr{\orgny -\unity/2 * \nr};
						\pgfmathsetmacro\nrr{\nq};
						\pgfmathsetmacro\unitxr{\unitx * \na};
						\pgfmathsetmacro\unityr{\unity * \na};
						
						\foreach \a in {1,...,\nrr}{
							\def\rootx{\orgnxr + \a * \unitxr/2};
							\def\rooty{\orgnyr -\unityr/2 * \a};
							\draw[ultra thick]  (\rootx, \rooty)
							-- (\rootx - \unitxr/2, \rooty + \unityr/2);
							\draw[ultra thick] (\rootx, \rooty)
							-- (\rootx + \a * \unitxr/2, \orgnyr);
						};
						
						\draw[ultra thick, loosely dotted, black]  (\orgnxr, \orgnyr) -- 
						(\orgnxr - \unitxr/3, \orgnyr + \unityr/3) ;
						
						\foreach \q in {0,...,\nrq}{
							\foreach \a in {0,...,\nr}{
								\def\x{\orgnx + \q * \unitx * \na + \a * \unitx};
								\def\y{\orgny};
								\ifthenelse{\q = 0 \AND \a = 3}{
									\filldraw[black] (\x, \y) circle (0pt) node[anyonnode]{\Huge$a^{m}_{q}$}; 
								}{
									\ifthenelse{\q = 1 \AND \a = 0}{
										\filldraw[black] (\x, \y) circle (0pt) node[anyonnode]{\Huge$a^{m+1}_0$}; 
									}{
										\ifthenelse{\q=1}{
											\ifthenelse{\a=3}{
												\filldraw[black] (\x, \y) circle (0pt) node[anyonnode]{\Huge$a^{m+1}_{q}$};
											}{
												\ifthenelse{\a=2}{
													\filldraw[black] (\x, \y) circle (0pt) node[anyonnode]{\Huge$a^{m+1}_{q-1}$};
												}{
													\filldraw[black] (\x, \y) circle (0pt) node[anyonnode]{\Huge$a^{m+1}_{\a}$};
												};
											};
										}{
											\ifthenelse{\a=2}{
												\filldraw[black] (\x, \y) circle (0pt) node[anyonnode]{\Huge $a^m_{q-1}$};
											}{
												\filldraw[black] (\x, \y) circle (0pt) node[anyonnode]{\Huge $a^m_{\a}$};
											};
										};
									};
								};
							};
						};
						
						\foreach \q in {1,...,\nq}{
							\foreach \a in {1,...,\nr}{
								\pgfmathsetmacro\p{\q - 1};
								\def\rootx{\orgnx + \p * \unitx * \na + \a * \unitx/2};
								\def\rooty{\orgny -\unity/2 * \a};
								\ifthenelse{\a=\nr}{
									\ifthenelse{\q=\nq}{
										\filldraw[black] (\rootx-\unitx/\den, \rooty) circle (0pt) node[anchor=east]{\Huge $i_{(m+1)q}$};
									}{
										\filldraw[black] (\rootx-\unitx/\den, \rooty) circle (0pt) node[anchor=east]{\Huge $i_{mq}$};
									}
								}{
									\ifthenelse{\a=2}{	
										\ifthenelse{\q=\nq}{
											\filldraw[black] (\rootx-\unitx/\den, \rooty) circle (0pt) node[anchor=east]{\Huge $i_{(m+1)(q-1)}$};
										}{
											\filldraw[black] (\rootx-\unitx/\den, \rooty) circle (0pt) node[anchor=east]{\Huge $i_{m(q-1)}$};
										}
									}{
										\ifthenelse{\a=1}{	
											\ifthenelse{\q=\nq}{
												\filldraw[black] (\rootx-\unitx/\den, \rooty) circle (0pt) node[anchor=east]{\Huge $i_{(m+1)1}$};
											}{
												\filldraw[black] (\rootx-\unitx/\den, \rooty) circle (0pt) node[anchor=east]{\Huge $i_{m1}$};
											}
								}}}
							};
						};
						
						\foreach \a in {0,...,\nrr}{
							\def\rootx{\orgnxr + \a * \unitxr/2};
							\def\rooty{\orgnyr -\unityr/2 * \a};
							\ifthenelse{\a=\nrr}{
								\filldraw[black] (\rootx, \rooty) circle (0pt) node[roundnode]{\Huge $j_{m}$};
							}{
								\ifthenelse{\a=0}{
									\filldraw[black] (\rootx, \rooty) circle (0pt) node[roundnode]{\Huge $j_{m-2}$};
								}{
									\filldraw[black] (\rootx, \rooty) circle (0pt) node[roundnode]{\Huge $j_{m-1}$};
								}
							}
						};
					} 
				} 
			} 
			
			\def\Kstateb{
				\resizebox{\versize}{!}{
					\tikz[baseline=.1ex,
					roundnode/.style={circle, draw=blue!60, fill=blue!5, minimum size=\minsize mm},
					anyonnode/.style={circle, draw=red!60, fill=red!5, thick, minimum size=\ratio*\minsize mm},
					]{
						\def\orgnx{0};
						\def\orgny{0};
						
						\pgfmathsetmacro\nr{\na-1};
						\pgfmathsetmacro\nrq{\nq-1};
						
						\foreach \q in {0,...,\nrq}{
							\foreach \a in {1,...,\nr}{
								\def\rootx{\orgnx + \q * \unitx * \na + \a * \unitx/2};
								\def\rooty{\orgny -\unity/2 * \a};
								\pgfmathsetmacro\temp{\nq-2};
								\ifthenelse{\a=2}{
									\draw[ultra thick, loosely dotted, black]  (\rootx, \rooty) -- 
									(\rootx - \unitx/2, \rooty + \unity/2) ;
								}{
									\draw[ultra thick]  (\rootx, \rooty) -- 
									(\rootx - \unitx/2, \rooty + \unity/2) ;
								}
								\draw[ultra thick] (\rootx, \rooty)
								-- (\rootx + \a * \unitx/2, \orgny);
							};
						};
						
						\pgfmathsetmacro\orgnxr{\orgnx -  \unitx * \na + \nr * \unitx/2};
						\pgfmathsetmacro\orgnyr{\orgny -\unity/2 * \nr};
						\pgfmathsetmacro\nrr{\nq};
						\pgfmathsetmacro\unitxr{\unitx * \na};
						\pgfmathsetmacro\unityr{\unity * \na};
						
						\foreach \a in {1,...,\nrr}{
							\def\rootx{\orgnxr + \a * \unitxr/2};
							\def\rooty{\orgnyr -\unityr/2 * \a};
							\draw[ultra thick]  (\rootx, \rooty)
							-- (\rootx - \unitxr/2, \rooty + \unityr/2);
							\draw[ultra thick] (\rootx, \rooty)
							-- (\rootx + \a * \unitxr/2, \orgnyr);
						};
						
						\draw[ultra thick, loosely dotted, black]  (\orgnxr, \orgnyr) -- 
						(\orgnxr - \unitxr/3, \orgnyr + \unityr/3) ;
						
						\def\xa{\orgnx + 3 * \unitx};
						\def\ya{\orgny};
						\def\xb{\orgnx + \unitx + 3 * \unitx};
						\def\yb{\orgny};
						\draw (\xb,\yb) .. controls (\xb, \yb+\unity/2) and (\xa,\ya+\unity/2) .. (\xa,\ya+\unity);
						\fill[white] (\xa+\unitx/2,\ya+\unity/2) circle (\unitx/10);
						\draw (\xa,\ya) .. controls (\xa,\ya+\unity/2) and (\xb,\yb+\unity/2) .. (\xb,\yb+\unity);
						
						\foreach \q in {0,...,\nrq}{
							\foreach \a in {0,...,\nr}{
								\def\x{\orgnx + \q * \unitx * \na + \a * \unitx};
								\def\y{\orgny};
								\ifthenelse{\q = 0 \AND \a = 3}{
									\filldraw[black] (\x, \y+\unity) circle (0pt) node[anyonnode]{\Huge $a^{m+1}_{0}$}; 
								}{
									\ifthenelse{\q = 1 \AND \a = 0}{
										\filldraw[black] (\x, \y+\unity) circle (0pt) node[anyonnode]{\Huge $a^m_q$}; 
									}{
										\ifthenelse{\q=1}{
											\ifthenelse{\a=3}{
												\filldraw[black] (\x, \y) circle (0pt) node[anyonnode]{\Huge $a^{m+1}_{q}$};
											}{
												\ifthenelse{\a=2}{
													\filldraw[black] (\x, \y) circle (0pt) node[anyonnode]{\Huge $a^{m+1}_{q-1}$};
												}{
													\filldraw[black] (\x, \y) circle (0pt) node[anyonnode]{\Huge $a^{m+1}_{\a}$};
												};
											};
										}{
											\ifthenelse{\a=2}{
												\filldraw[black] (\x, \y) circle (0pt) node[anyonnode]{\Huge $a^m_{q-1}$};
											}{
												\filldraw[black] (\x, \y) circle (0pt) node[anyonnode]{\Huge $a^m_{\a}$};
											};
										};
									};
								};
							};
						};
						
						\foreach \q in {1,...,\nq}{
							\foreach \a in {1,...,\nr}{
								\pgfmathsetmacro\p{\q - 1};
								\def\rootx{\orgnx + \p * \unitx * \na + \a * \unitx/2};
								\def\rooty{\orgny -\unity/2 * \a};
								\ifthenelse{\a=\nr}{
									\ifthenelse{\q=\nq}{
										\filldraw[black] (\rootx-\unitx/\den, \rooty) circle (0pt) node[anchor=east]{\Huge $i_{(m+1)q}$};
									}{
										\filldraw[black] (\rootx-\unitx/\den, \rooty) circle (0pt) node[anchor=east]{\Huge $i_{mq}$};
									}
								}{
									\ifthenelse{\a=2}{	
										\ifthenelse{\q=\nq}{
											\filldraw[black] (\rootx-\unitx/\den, \rooty) circle (0pt) node[anchor=east]{\Huge $i_{(m+1)(q-1)}$};
										}{
											\filldraw[black] (\rootx-\unitx/\den, \rooty) circle (0pt) node[anchor=east]{\Huge $i_{m(q-1)}$};
										}
									}{
										\ifthenelse{\a=1}{	
											\ifthenelse{\q=\nq}{
												\filldraw[black] (\rootx-\unitx/\den, \rooty) circle (0pt) node[anchor=east]{\Huge $i_{(m+1)1}$};
											}{
												\filldraw[black] (\rootx-\unitx/\den, \rooty) circle (0pt) node[anchor=east]{\Huge $i_{m1}$};
											}
								}}}
							};
						};
						
						\foreach \a in {0,...,\nrr}{
							\def\rootx{\orgnxr + \a * \unitxr/2};
							\def\rooty{\orgnyr -\unityr/2 * \a};
							\ifthenelse{\a=\nrr}{
								\filldraw[black] (\rootx, \rooty) circle (0pt) node[roundnode]{\Huge $j_{m}$};
							}{
								\ifthenelse{\a=0}{
									\filldraw[black] (\rootx, \rooty) circle (0pt) node[roundnode]{\Huge $j_{m-2}$};
								}{
									\filldraw[black] (\rootx, \rooty) circle (0pt) node[roundnode]{\Huge $j'_{m-1}$};
								}
							}
						};
					} 
				} 
			} 
			
			\def\Kstatef{
				\resizebox{\versize}{!}{
					\tikz[baseline=.1ex,
					roundnode/.style={circle, draw=blue!60, fill=blue!5, minimum size=\minsize mm},
					anyonnode/.style={circle, draw=red!60, fill=red!5, thick, minimum size=\ratio*\minsize mm},
					]{
						\def\orgnx{0};
						\def\orgny{0};
						
						\pgfmathsetmacro\nr{\na-1};
						\pgfmathsetmacro\nrq{\nq-1};
						
						\foreach \q in {0,...,\nrq}{
							\foreach \a in {1,...,\nr}{
								\def\rootx{\orgnx + \q * \unitx * \na + \a * \unitx/2};
								\def\rooty{\orgny -\unity/2 * \a};
								\pgfmathsetmacro\temp{\nq-2};
								\ifthenelse{\a=2}{
									\draw[ultra thick, loosely dotted, black]  (\rootx, \rooty) -- 
									(\rootx - \unitx/2, \rooty + \unity/2) ;
								}{
									\draw[ultra thick]  (\rootx, \rooty) -- 
									(\rootx - \unitx/2, \rooty + \unity/2) ;
								}
								\draw[ultra thick] (\rootx, \rooty)
								-- (\rootx + \a * \unitx/2, \orgny);
							};
						};
						
						\pgfmathsetmacro\orgnxr{\orgnx -  \unitx * \na + \nr * \unitx/2};
						\pgfmathsetmacro\orgnyr{\orgny -\unity/2 * \nr};
						\pgfmathsetmacro\nrr{\nq};
						\pgfmathsetmacro\unitxr{\unitx * \na};
						\pgfmathsetmacro\unityr{\unity * \na};
						
						\foreach \a in {1,...,\nrr}{
							\def\rootx{\orgnxr + \a * \unitxr/2};
							\def\rooty{\orgnyr -\unityr/2 * \a};
							\draw[ultra thick]  (\rootx, \rooty)
							-- (\rootx - \unitxr/2, \rooty + \unityr/2);
							\draw[ultra thick] (\rootx, \rooty)
							-- (\rootx + \a * \unitxr/2, \orgnyr);
						};
						
						\draw[ultra thick, loosely dotted, black]  (\orgnxr, \orgnyr) -- 
						(\orgnxr - \unitxr/3, \orgnyr + \unityr/3) ;
						
						\foreach \q in {0,...,\nrq}{
							\foreach \a in {0,...,\nr}{
								\def\x{\orgnx + \q * \unitx * \na + \a * \unitx};
								\def\y{\orgny};
								\ifthenelse{\q = 0 \AND \a = 3}{
									\filldraw[black] (\x, \y) circle (0pt) node[anyonnode]{\huge $a^{m+1}_{0}$}; 
								}{
									\ifthenelse{\q = 1 \AND \a = 0}{
										\filldraw[black] (\x, \y) circle (0pt) node[anyonnode]{\Huge $a^m_q$}; 
									}{
										\ifthenelse{\q=1}{
											\ifthenelse{\a=3}{
												\filldraw[black] (\x, \y) circle (0pt) node[anyonnode]{\Huge $a^{m+1}_{q}$};
											}{
												\ifthenelse{\a=2}{
													\filldraw[black] (\x, \y) circle (0pt) node[anyonnode]{\Huge $a^{m+1}_{q-1}$};
												}{
													\filldraw[black] (\x, \y) circle (0pt) node[anyonnode]{\Huge $a^{m+1}_{\a}$};
												};
											};
										}{
											\ifthenelse{\a=2}{
												\filldraw[black] (\x, \y) circle (0pt) node[anyonnode]{\Huge $a^m_{q-1}$};
											}{
												\filldraw[black] (\x, \y) circle (0pt) node[anyonnode]{\Huge $a^m_{\a}$};
											};
										};
									};
								};
							};
						};
						
						\foreach \q in {1,...,\nq}{
							\foreach \a in {1,...,\nr}{
								\pgfmathsetmacro\p{\q - 1};
								\def\rootx{\orgnx + \p * \unitx * \na + \a * \unitx/2};
								\def\rooty{\orgny -\unity/2 * \a};
								\ifthenelse{\a=\nr}{
									\ifthenelse{\q=\nq}{
										\filldraw[black] (\rootx-\unitx/\den, \rooty) circle (0pt) node[anchor=east]{\Huge $i'_{(m+1)q}$};
									}{
										\filldraw[black] (\rootx-\unitx/\den, \rooty) circle (0pt) node[anchor=east]{\Huge $i'_{mq}$};
									}
								}{
									\ifthenelse{\a=2}{	
										\ifthenelse{\q=\nq}{
											\filldraw[black] (\rootx-\unitx/\den, \rooty) circle (0pt) node[anchor=east]{\Huge $i'_{(m+1)(q-1)}$};
										}{
											\filldraw[black] (\rootx-\unitx/\den, \rooty) circle (0pt) node[anchor=east]{\Huge $i_{m(q-1)}$};
										}
									}{
										\ifthenelse{\a=1}{	
											\ifthenelse{\q=\nq}{
												\filldraw[black] (\rootx-\unitx/\den, \rooty) circle (0pt) node[anchor=east]{\Huge $i'_{(m+1)1}$};
											}{
												\filldraw[black] (\rootx-\unitx/\den, \rooty) circle (0pt) node[anchor=east]{\Huge $i_{m1}$};
											}
								}}}
							};
						};
						
						\foreach \a in {0,...,\nrr}{
							\def\rootx{\orgnxr + \a * \unitxr/2};
							\def\rooty{\orgnyr -\unityr/2 * \a};
							\ifthenelse{\a=\nrr}{
								\filldraw[black] (\rootx, \rooty) circle (0pt) node[roundnode]{\Huge $j_{m}$};
							}{
								\ifthenelse{\a=0}{
									\filldraw[black] (\rootx, \rooty) circle (0pt) node[roundnode]{\Huge $j_{m-2}$};
								}{
									\filldraw[black] (\rootx, \rooty) circle (0pt) node[roundnode]{\Huge $j'_{m-1}$};
								}
							}
						};
					} 
				} 
			} 

			\begin{table*}[h]
				\begin{minipage}{\linewidth}
					\begin{align}
						\label{eq:general-K}
						\sigma_{m\times N}
						\begin{array}{c}
							\Kstatei
						\end{array}
						= &
						\begin{array}{c}
							\Kstateb
						\end{array}
						\nonumber \\ 
						=\sum_{j'_{m-1} i'_{mq}i'_{(m+1)1)}\cdots i'_{(m+1)q}}
						\left(K^{j_{m-2} j_{m}}_{i_{m(q-1)}} \right)^{j_{m-1} i_{mq} \cdots i_{(m+1)q}}_{{j'}_{m-1} {i'}_{mq}\cdots i'_{(m+1)q}}
						&
						\begin{array}{c}
							\Kstatef
						\end{array}
					\end{align}
					such that:
					\begin{align}
						\label{eq:general-K-definition}
						& \left(K^{j_{m-2} j_{m}}_{i_{m(q-1)}} \right)^{j_{m-1} i_{mq} \cdots i_{(m+1)q}}_{{j'}_{m-1} {i'}_{mq}\cdots i'_{(m+1)q}} = 
						\nonumber\\&
						\sum_{k}
						\left(F_{j_{m-2} i_{mq} i_{(m+1)q}}^{j_m} \right)^{j_{m-1}}_k 
						\left(L^{k}_{i_{m(q-1)}}\right)^{i_{mq} \cdots i_{(m+1)q}}_{i'_{mq} \cdots i'_{(m+1)q}}
						\left(F_{j_{m-2} i'_{mq} i'_{(m+1)q}}^{\dagger j_m}\right)_{j'_{m-1}}^k.
					\end{align}
				\end{minipage}
			\end{table*}
			
		\end{document}